\documentclass[manuscript,screen]{acmart}

\AtBeginDocument{%
  \providecommand\BibTeX{{%
    \normalfont B\kern-0.5em{\scshape i\kern-0.25em b}\kern-0.8em\TeX}}}

\setcopyright{acmcopyright}
\copyrightyear{2018}
\acmYear{2018}
\acmDOI{XXXXXXX.XXXXXXX}

\acmConference[Conference acronym 'XX]{Make sure to enter the correct
  conference title from your rights confirmation emai}{June 03--05,
  2018}{Woodstock, NY}
\acmPrice{15.00}
\acmISBN{978-1-4503-XXXX-X/18/06}

\usepackage[utf8]{inputenc} 
\usepackage[T1]{fontenc}    
\usepackage{hyperref}       
\usepackage{url}            
\usepackage{booktabs}       
\usepackage{amsfonts}       
\usepackage{nicefrac}       
\usepackage{microtype}      
\usepackage{xcolor}         
\usepackage{caption}
\usepackage{subcaption}

\usepackage{hhline}
\usepackage{multirow}
\usepackage{amsthm}

\newtheorem{theorem}{Theorem}[section]
\newtheorem{proposition}[theorem]{Proposition}

\newcommand{\ict}{i,t}
\newcommand{\mutilde}{\tilde{\mu}}
\newcommand{\gammatilde}{\tilde{\gamma}}
\newcommand{\ctilde}{\tilde{c}}
\newcommand{\muprmtilde}{\tilde{\mu^\prime}}
\newcommand{\mutildeit}{\mutilde_{\ict}}
\newcommand{\muprmtildeit}{\muprmtilde_{\ict}}
\newcommand{\mit}{m_{\ict}}
\newcommand{\nuict}{\nu_{\ict}}
\newcommand{\mubarit}{\bar{\mu}_{\ict}}
\newcommand{\nubarit}{\bar{\nu}_{\ict}}
\newcommand{\As}{A_s}
\newcommand{\Ad}{A_d}
\newcommand{\Aprms}{A^\prime_s}
\newcommand{\Aprmd}{A^\prime_d}
\newcommand{\calN}{\mathcal{N}}
\newcommand{\calG}{\mathcal{G}}
\newcommand{\calV}{\mathcal{V}}
\newcommand{\calE}{\mathcal{E}}
\newcommand{\calT}{\mathcal{T}}
\newcommand{\Rmbb}{\mathbb{R}}
\newcommand{\Embb}{\mathbb{E}}
\newcommand{\tautotal}{\tau_{\text{total}}}
\newcommand{\tauglobal}{\tau_{\text{vol}}}
\newcommand{\taucond}{\tau_{\text{dist}}}
\newcommand{\hattautotal}{\hat\tau_{\text{total}}}
\newcommand{\hattauglobal}{\hat\tau_{\text{vol}}}
\newcommand{\hattaucond}{\hat\tau_{\text{dist}}}
\newcommand{\vshift}{\vec v_{\text{shift}}}
\DeclareMathOperator{\proj}{proj}

\begin{document}

\title{A Unified Representation Framework for Rideshare Marketplace Equilibrium and Efficiency}


\author{Alex Chin}
\affiliation{%
  \institution{Lyft}
  \city{San Francisco, CA}
  \country{USA}}

\author{Zhiwei (Tony) Qin}
\affiliation{%
  \institution{Lyft}
  \city{San Francisco, CA}
  \country{USA}}







\begin{abstract}
    Ridesharing platforms are a type of two-sided marketplace where ``supply-demand balance'' is critical for market efficiency and yet is complex to define and analyze. We present a unified analytical framework based on the graph-based equilibrium metric (GEM) for quantifying the supply-demand spatiotemporal state and efficiency of a ridesharing marketplace. GEM was developed as a generalized Wasserstein distance between the supply and demand distributions in a ridesharing market and has been used as an evaluation metric for algorithms expected to improve supply-demand alignment. Building upon GEM, we develop SD-GEM, a dual-perspective (supply- and demand-side) representation of rideshare market equilibrium. We show that there are often disparities between the two views and examine how this dual-view leads to the notion of market efficiency, in which we propose novel statistical tests for capturing improvement and explaining the underlying driving factors. 

\end{abstract}






\maketitle

\section{Introduction}\label{sec:intro}
A ridesharing platform connects riders with transportation needs to drivers open for work. Both riders and drivers are important to the platform --- more rider demand means that drivers on the platform enjoy higher utilization and earning rate, and more driver supply means that riders wait for a shorter period before being matched to a driver and subsequently picked up. The demand and supply on a ridesharing platform are typically matched in a batched manner --- they are grouped in time windows of equal size and matched within a certain dispatch radius. After transporting passengers to their destinations, drivers become available again for new requests. While drivers are idle, repositioning strategies guide their distribution to better match that of the future demand for higher market efficiency. Pricing also shapes the demand distribution, directly impacting rider request conversion. In addition, rider and driver incentives leverage monetary rewards to grow these two participant populations and to incentivize their spatiotemporal distributions to match the platform's needs. For detailed  descriptions of the various components and the associated problems of a ridesharing platform, we refer the readers to the surveys \cite{wang2019ridesourcing,qin2022reinforcement}. To efficiently operate a ridesharing marketplace, it is important to be able to predict the future supply and demand at different granularities and over different time horizons. Machine learning, and deep learning in particular, has been the main methodological tool for supply-demand (SD) forecasting \cite{tong2017simpler,yao2018deep,geng2019spatiotemporal}.

Simple questions like “Is there enough driver supply on the platform?” may lead to complicated answers, because within the context of a two-sided marketplace we have to be more specific. Are we asking 1) “are there enough drivers at the right place to maintain a good service level”? Or 2) “are there enough drivers assuming the drivers are willing to accept a request no matter how far it is?” Or 3) “are there the right number of drivers online to maintain an attractive earning rate?” The presumed context leads to different answers, and that is the interesting and challenging part of a two-sided marketplace.

The graph-based equilibrium metric (GEM) \cite{zhou2021graph}
was developed as a generalized Wasserstein distance between the supply and demand distributions in a ridesharing market and has been used as an evaluation metric for algorithms intervening the SD alignment. (See \cite{zhou2021graph} and the references therein for other types of distributional distances.) Its computation is based on solving an asymmetric optimal transport problem that resembles the rideshare dispatch problem. The original GEM's single state perspective, however, makes it insufficient to answer the questions posed above precisely or to form a notion of market efficiency in utilizing the supply for demand fulfillment because it is unable to represent any potential trade-off between the supply- and demand-side states. In this paper,  we argue that a two-sided view of the market state is needed for a complete representation of market equilibrium and to test for efficiency change. 

There are extensive works in the ridesharing literature on marketplace optimization to improve a rideshare platform's performance, typically in market equilibrium and efficiency. Some recent works cover algorithms for dispatch \cite{xu2018large,tang2019deep,tang2021value,eshkevari2022reinforcement,han2022real,zhou2019multi}, repositioning \cite{jtq2021repos,chaudhari2020learn,lin2018efficient,ong2021driver,braverman2019empty}, pricing \cite{ma2022spatio,chen2021spatial,hu2021surge,bimpikis2019spatial,yan2020dynamic}, and incentives \cite{wu2022spatio,shang2019environment}. There have also been works focusing on investigations and analyses of the driving factors for rideshare market efficiency \cite{ke2020pricing,ghili2021spatial,han2017quantification,al2021drives}.

There are a wide spectrum of observational marketplace metrics adopted from the rider, driver, and platform perspectives, each of which measures a particular aspect of the marketplace performance. On the rider side, many works look at the average pick-up expected-time-to-arrival (ETA) to the request origin, pre-match / post-match cancellation rates, and request fulfillment rate. The driver-side metrics focus on driver online hours, utilization, and earning rate. Finally, the platform metrics typically measure total bookings, rides, and incentive spend efficiency. 

A major challenge brought by the long list of metrics is the distraction and noise to the analyst and decision-maker when the numbers send mixed messages, making it hard to grasp what is really happening at the core. In addition, these metrics are often heterogeneous in the unit of measurement, and making quantitative trade-offs among them is not straightforward. There is a strong need of a fundamental representation of the SD state to succinctly explain the variations in the basket of marketplace metrics. There also currently lacks a unified notion of market efficiency for ridesharing and how to perform statistical tests for it.

Our contributions and the organization of the paper are as follows. In Section~\ref{sec:gem}, we review the GEM framework and extend it in Section~\ref{sec:gem_duals} to include dual-based market state diagnostic tools. In Section \ref{sec:dual_view} we present SD-GEM, a two-sided-view framework for a comprehensive representation of rideshare market equilibrium states, and in Section \ref{sec:efficiency} we develop a unified notion and quantification of marketplace efficiency within this framework. Section \ref{sec:estimation} develops novel statistical tests for measuring change in efficiency, Section \ref{sec:applications} offers examples and validation based on real-world experiment data, and Section~\ref{sec:conclusion} concludes.



\section{GEM Overview}\label{sec:gem}
We review the GEM framework as described in~\citep{zhou2021graph}. We first partition the geographic space into $N$ disjoint areas. Let $\calG = (\calV, \calE)$ be a (possibly weighted and directed) graph, for which each of the vertices $v \in \calV$ is associated with the $N$ areas so that $|\calV| = N$. Each edge $e_{ij} = (v_i, v_j)$ is associated with a transportation cost $w_{ij}$ for traveling from area $i$ to area $j$. From $\calG$ and the weight matrix $W = (w_{ij})$ we can define two additional quantities: the neighborhood $\calN_i$ and the cost matrix $C = (c_{ij})$. The neighborhood $\calN_i$ is the set of locations $j$ for which it is possible for a supply unit to be transported from location $i$. The cost matrix is derived from the edge weights and is taken to be $c_{ij} = \inf_{\text{paths }p} \sum_{(k,\ell) \in p} w_{k\ell}$, the transportation cost of the cheapest path from $i$ to $j$.
\subsection{Optimization}

Following the notation in~\citep{zhou2021graph}, we let $\mu$ and $\nu$ be discrete measures on $\calV$ representing the observed supply and demand in each cell, respectively. For example, $\mu_i$ could be the number of drivers in cell $i$ available to be dispatched, and $\nu_i$ could be the number of open ride requests in cell $i$.

The GEM optimization problem is
\begin{equation}
\label{eqn:objective}
\rho_\lambda (\mu, \nu) = \min_{\gamma \in \Gamma} \left\{\|\nu - \tilde \mu\|_1 + \lambda \sum_{v_i \in \calV} \sum_{v_j \in \calV} c_{ij} \gamma_{ij}\right\}
\end{equation}
subject to
\[\sum_{v_j \in \calN_i} \gamma_{ij} = \mu_i, \quad \sum_{v_j \not \in \calN_i} \gamma_{ij} = 0, \quad \tilde \mu_i = \sum_{v_i \in \calN_j} \gamma_{ji}, \quad \text{ for all } v_i \in \calV.\]
The decision variable is the transport plan matrix $(\gamma_{ij})$, where $\gamma_{ij}$ is the amount of supply transported from location $i$ to location $j$. $\tilde \mu$ is fully determined as a function of $\gamma$, and represents the optimal supply distribution after transporting $\mu$ via plan $\gamma$. $\mutilde$ offers a more realistic view of supply availability for each cell than the observed supply $\mu$ because it has accounted for the dispatch radius (encoded by $\calN_i$) and its effect on the perceived supply distribution. The objective value $\rho_{\lambda}$ is a generalized (asymmetric) Wasserstein distance and a semi-metric, as shown in \cite{zhou2021graph}.

\subsection{Local and Global Balance Measures}
Given the solution of~\eqref{eqn:objective}, we define the \emph{local supply-demand gap} as
\[m_i = \log \tilde \mu_i - \log \nu_i,\]
where zero supply or demand quantities are cushioned by $\epsilon > 0$.
This is similar to the quantities $\operatorname{DSr}_i = \nu_i / \max(\tilde \mu_i, 1)$ and $\operatorname{DSd}_i = \nu_i  - \tilde \mu_i$ proposed by~\citep{zhou2021graph}, but we find $m_i$ as defined above to be better behaved in practice, as it is both additive and market volume invariant. A healthy equilibrium state for the marketplace corresponds to values of the SD gap close to zero.

Now, consider solving \eqref{eqn:objective} across multiple units of time $t$, producing a matrix of SD gaps $m_{i,t}$. Denote similarly the quantities $\nu_{i,t}, \mu_{i,t}$ and $\tilde \mu_{i,t}$. \citep{zhou2021graph} suggest aggregating the $m_{i,t}$ up to a coarser unit of granularity, which yields a global view of the market balance state:
\begin{equation}\label{eq:demand_view}
    A_d = \frac{\sum_{i, t} m_{i,t} \nu_{i,t}}{\sum_{i,t} \nu_{i,t}}=\Embb_{(\ict)\sim\nu}[\mit].
\end{equation}
A positive value of $\Ad$ indicates over-supply, and a negative value indicates under-supply.
The summation can be considered over any aggregate units of time and space that are of interest. The choice of $\nu_{i,t}$ as the weight coefficient highlights areas with high demand, and thus $A_d$ provides a \emph{demand-centric view} of the marketplace equilibrium state\footnote{We also interchangeably call $\Ad$ the demand-centric SD \emph{index}.}, which is the expectation of the local SD gap with respect to the demand spatiotemporal distribution. It represents the expected local market state that a randomly chosen rider sees, and it answers Question 1) in Section \ref{sec:intro}. A highly negative $\Ad$ has negative implication to the rider-side metrics---long pre-match waiting time or low request fulfillment rate. In this paper, we argue that varying the choice of weight (or equivalently, the distribution with which the expectation is computed) yields a more complete perspective that can lead to actionable policy choices.

\section{Market State Diagnostics}\label{sec:gem_duals}
We extend the original GEM framework to develop rideshare market state diagnostics based on the dual solution to \eqref{eqn:objective}.
Let us define vector $\gammatilde := \{\gamma_{ij}, j \in \calN_i\}\in\Rmbb^{N_0}$, where $N_0=\sum_1^N n_i$. $\ctilde\in\Rmbb^{N_0}$ is the vector containing all $\{\gamma_{ij}\in\gammatilde\}$.  
GEM \eqref{eqn:objective} can be written as the following linear program (LP) and can be solved by an LP solver \cite{zhou2021graph}:

\begin{align}\label{eq:gem_lp}
    \min_{\gammatilde\geq 0,s\geq 0} \;&\; \mathbf{1}^Ts + \lambda\ctilde^T\gammatilde \\
     s.t. \;&\; A_1\gammatilde=\mu, \label{eq:gem_lp_supply}\\
        \;&\; A_2\gammatilde+s\geq\nu, \label{eq:gem_lp_under}\\
        \;&\; A_2\gammatilde-s \leq \nu, \label{eq:gem_lp_over}.
\end{align}

We denote the dual variables corresponding to the three constraints in \eqref{eq:gem_lp_supply}-\eqref{eq:gem_lp_over} by $w$, $u^u$, and $u^o$, respectively. The pair of constraints \eqref{eq:gem_lp_under} and \eqref{eq:gem_lp_over} correspond to the $L^1$-norm in \eqref{eqn:objective}. The constraint \eqref{eq:gem_lp_under} represents under-supplied situations, with a non-negative slack w.r.t. the demand, whereas \eqref{eq:gem_lp_supply} represents over-supplied situations, with a non-positive slack. Hence, we have for the demand duals, $u^u \geq 0$, and $u^o \leq 0$. From complementary slackness conditions, we know that for a given spatiotemporal position $(\ict)$, if after solving the LP \eqref{eq:gem_lp}, $(A_2\gammatilde)_{\ict} > \nu_{\ict}$, then $u^u_{\ict} = 0$, and if $(A_2\gammatilde)_{\ict} < \nu_{\ict}$, then $u^o_{\ict} = 0$. In all cases, we look at $u := u^u + u^o$ for the marginal effect of adding (or removing) one request at position $(\ict)$ to the GEM objective $\rho$ because doing so will perturb the right-hand-side of both constraints \eqref{eq:gem_lp_under} and \eqref{eq:gem_lp_over} by one positive unit. Note that the duals convey the marginal values in system SD gap and transport cost of the perturbations, whereas $\mit$ represents the local SD gap only.

\begin{figure}
    \centering
    \includegraphics[width=0.9\textwidth]{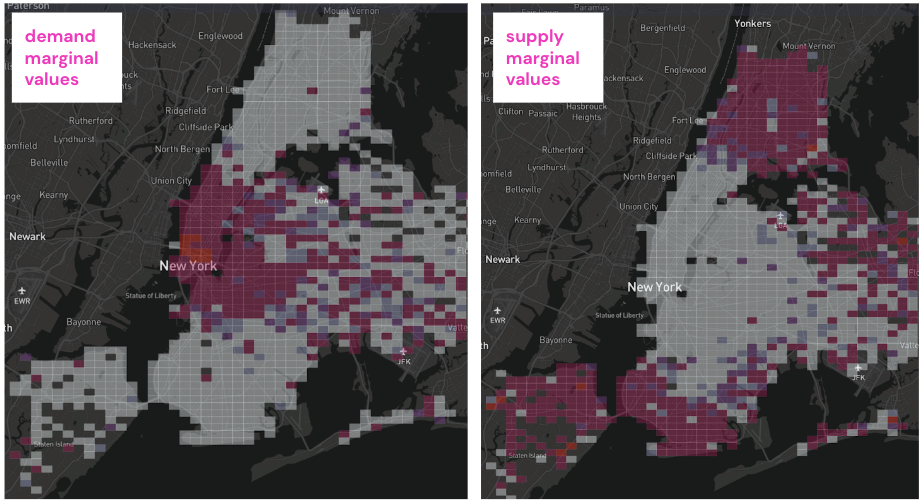}
    \caption{A random historical snapshot of the distributions of the GEM duals. More reddish color: higher (positive) dual values. Lighter grey color: lower (negative) dual values. }
    \label{fig:mevs}
\end{figure}

The GEM demand duals provide convenient diagnostic analytics for demand shaping strategies (e.g., \cite{wu2022spatio,chen2021spatial,hu2021surge}) to maintain a healthy market balance when cross-referencing them. A negative $u_{\ict}$ at a particular location and time, for example, means that demand could be increased because an incremental request would improve the GEM objective value. On the other hand, a positive $u_{\ict}$ means that the market balance may even improve with lower demand in the spatiotemporal zone. The left-hand plot of Figure \ref{fig:mevs} shows  visualization of the demand duals distribution.

The supply dual $w_{\ict}$ represents the marginal effect on $\rho$ of adding one driver to the $(\ict)$ position.  The $w$ variables can offer guiding analytics for supply positioning strategies (e.g., \cite{ong2021driver,jtq2021repos,lin2018efficient}) because they indicate where drivers are needed the most and where we may reposition drivers from. The right-hand plot of Figure \ref{fig:mevs} shows visualization of the supply duals distribution. We see that it corresponds well to the distribution of the demand duals, which makes intuitive sense - the areas where we need drivers the most are generally the areas that we may have too much demand. 

\section{Two-sided View of Market Balance}\label{sec:dual_view} 
Now, we present SD-GEM, our proposed two-sided-view framework built upon the original GEM. SD-GEM are designed to allow us to easily get a more complete picture of the market state, explaining both rider and driver-side observations. In particular, the level of disparity of the two views has important actionable insights, and we develop principled quantification of market efficiency change through the shift of a market's state within this two-sided view (Sections \ref{sec:efficiency}-\ref{sec:applications}).

\subsection{Supply-centric View}\label{sec:supply_view}
In SD-GEM, we compute the \emph{supply-centric view} of the market equilibrium in addition to the existing demand-centric view: 
\begin{equation}\label{eq:supply_view}
    A_s := \frac{\sum_{\ict}\mit\mutildeit}{\sum_{\ict}\mutildeit}=\Embb_{(\ict)\sim\mutilde}[\mit]
\end{equation}
$\As$ has the same definitions for its signs as $\Ad$ \eqref{eq:demand_view}, but $\As$ puts more weights to the SD gap $\mit$ of the areas where supplies are concentrated, since the expectation is with respect to the distribution of the dispatched supply. It represents the average perception of the local market balance state from the view of a randomly sampled driver.
The supply-centric view answers Question 3) in Section \ref{sec:intro} and drives supply-side observational metrics like driver utilization and earning rate. A highly positive $\As$ means that more drivers in general have to compete for the limited demand, leading to low utilization and earning rate.

\paragraph{Remark}
We note the similarity between the supply/demand-centric views and the forward/reverse \emph{Kullback-Leibler (KL) Divergence}. We can write \eqref{eq:supply_view} as $\As = \sum_{\ict\in\calT}\mubarit\left(\log(\frac{\mubarit}{\nubarit})+\log(\frac{M}{N})\right)$, where $\calT$ is the spatiotemporal interval of aggregation, $M=\sum_{\ict}\mutildeit$, and $N=\sum_{\ict}\nu_{\ict}$. The normalized supply and demand, $\mubarit=\frac{\mutildeit}{M}$ and $\nubarit=\frac{\nu_{\ict}}{N}$, are the probability distributions of  $\mutilde$ and $\nu$.  When the total volumes of the supply and demand are equal, i.e., $M=N$, we have $\As = D_{KL}(\bar{\mu}||\bar{\nu})$, the (forward) KL divergence between the probability distributions $\bar{\mu}$ and $\bar{\nu}$. Similarly, under the same condition, we have $\Ad = -D_{KL}(\bar{\nu}||\bar{\mu})$, the negative of the reverse KL divergence. This connection makes it easy to see that our two-sided views of market balance measure not only the distributional alignment but also the gap between the SD volumes, since the $\log(\frac{M}{N})$ term is non-zero in general. It also explains why it is insufficient to consider the total SD volumes in a market for SD equilibrium. Similar to the forward and reverse KL divergences, the two-sided views $\Ad$ and $\As$ are not equal in general. In the subsequent sections, we will show how these two values capture the core of a rideshare marketplace state and how to develop statistical tests to measure changes in market efficiency based on movements in their 2D space.

\subsection{Illustrative Examples}\label{sec:examples}
We use a set of stylized dispatch examples to illustrate the two-sided views introduced above. In each example, the pink circles represent drivers, and the blue triangles represent requests. The entire region is divided into four geographical cells.
Drivers can be matched to only requests in their current or adjacent cells (not diagonally). Black arrows represent some non-obvious matches.
The pairs of numbers around the grid represent $\mutilde : \nu$  (optimal dispatched supply vs demand), and $\epsilon = 0.5$.

\begin{figure}
    \centering
    \begin{subfigure}[b]{0.3\linewidth}
        \centering
        \includegraphics[width=\textwidth]{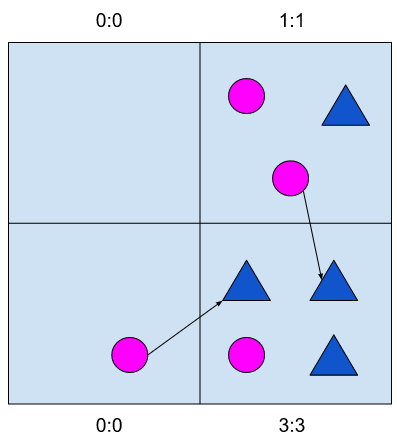}
        \caption{}
        \label{subfig:globally_balanced}
    \end{subfigure}
    \hspace{0.15\linewidth}
    \begin{subfigure}[b]{0.3\linewidth}
        \centering
        \includegraphics[width=\textwidth]{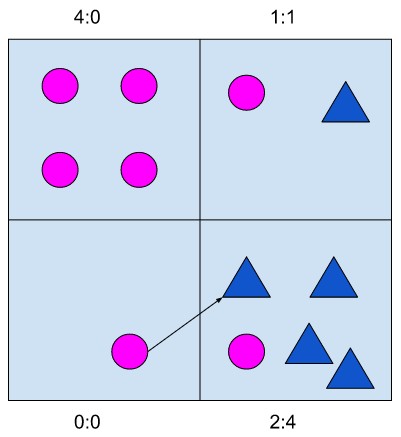}
        \caption{}
        \label{subfig:sd_misaligned}
    \end{subfigure}
    \begin{subfigure}[b]{0.3\linewidth}
        \centering
        \includegraphics[width=\textwidth]{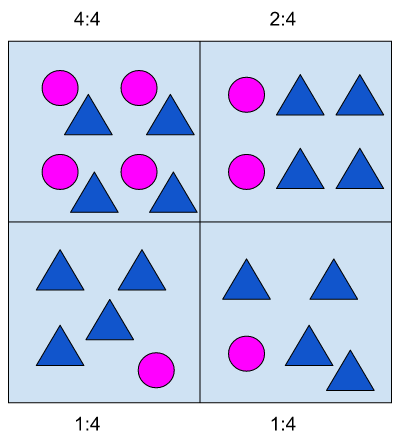}
        \caption{}
        \label{subfig:globally_undersupplied}
    \end{subfigure}
    \hspace{0.15\linewidth}
    \begin{subfigure}[b]{0.3\linewidth}
        \centering
        \includegraphics[width=\textwidth]{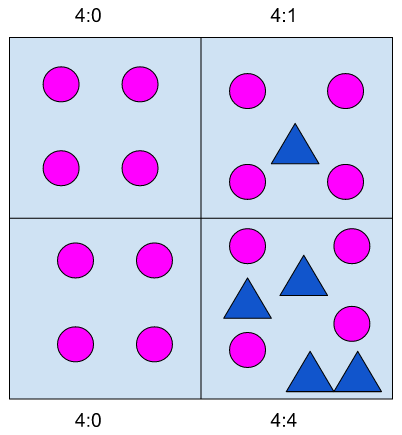}
        \caption{}
        \label{subfig:globally_oversupplied}
    \end{subfigure}
    \caption{Illustrative examples of the two-sided views of SD balance. (a) Globally balanced (b) SD misaligned (c) Globally under-supplied (d) Globally over-supplied}
    \label{fig:examples}
\end{figure}

\paragraph{Globally balanced (Figure \ref{subfig:globally_balanced})}
Both riders and drivers see the same balanced state: $\Ad=0, \; \As=0$. This is generally an ideal state.

\paragraph{SD misaligned (Figure \ref{subfig:sd_misaligned})}
Riders see a generally under-supplied state, while drivers see a generally over-supplied state:\\ $\Ad=-0.24,\; \As=1.05$. (The detailed calculation is omitted since the verification is straightforward.)

The two views of SD-GEM present SD indices of opposite signs. This indicates that some drivers are not at the right locations (within the required time slice) to be matched to requests, and there are opportunities for repositioning the drivers to better align with the demand distribution.

\paragraph{Globally under-supplied (Figure \ref{subfig:globally_undersupplied})}
Both riders and drivers see a generally under-supplied state: $A_s=-0.87,\; A_d=-0.52$. Driver acquisition and retention are critical in this state regime. 

\paragraph{Globally over-supplied (Figure \ref{subfig:globally_oversupplied})}
Both riders and drivers see a generally over-supplied state: $A_s=0.28, \;A_d=1.39$. Demand acquisition and/or supply control are important in this state regime.

\subsection{State Classification}
\begin{figure}
    \centering
    \includegraphics[width=0.6\textwidth]{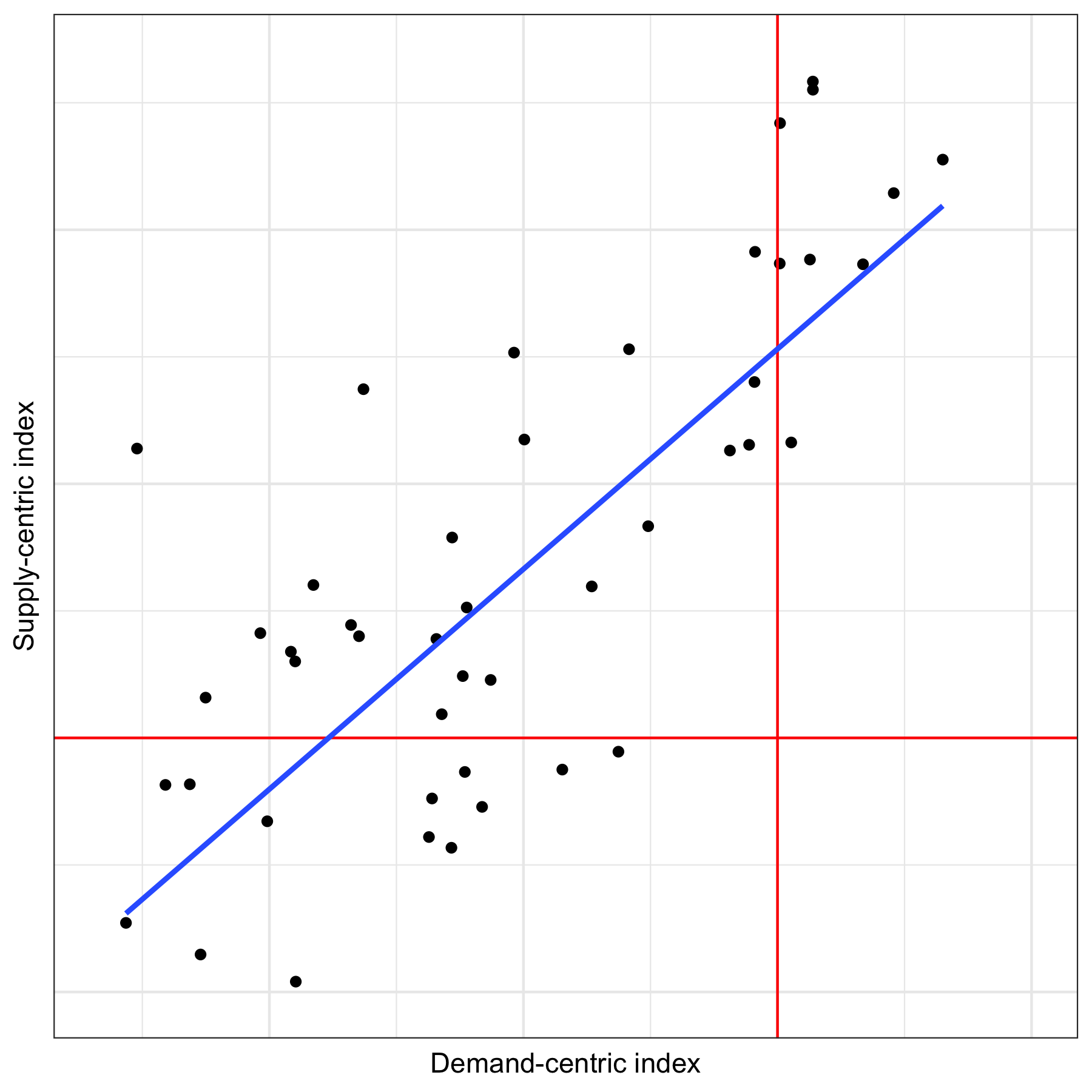}
    \caption{Scatter plot of the supply- and demand- centric views for a sample of rideshare markets over a one-month period during the pandemic. The red lines indicate the $x=0$ and $y=0$ intercepts, and the blue line is the OLS regression line.}
    \label{fig:global_dual_view}
\end{figure}

The four scenarios from Section \ref{sec:examples} provide a good classification of the market balance state. In general, the state of any city falls into the following nine-cell matrix in Table \ref{tab:state_classification}, representing a partition of the 2D space spanned by $(A_s, A_d)$. The most desirable state is that both $\As$ and $\Ad$ are close to zero, which corresponds to consistent perceptions of good SD balance from both the driver and rider perspectives. (See Section \ref{sec:queuing} for further explanations on this.) To determine what should be considered `close to zero' (or the intervals on the $\As$ and $\Ad$ axes defining the middle column and row in the nine-cell matrix, i.e., the neutral intervals), we took an empirical approach. We analyzed the two indices against common rider and driver-side observational metrics, such as pick-up ETA, rider cancellation, and driver utilization, for a large number of (city, time period) combinations. The empirical evidence suggests that the boundaries for the desirable state (beyond which the relevant observational metrics generally start to deteriorate) are in the form of [-$\delta$, 0] for $\Ad$ and [0, $\delta$] for $\As$, where we set $\delta=0.2$. Note that these analyses are time-invariant (or market-state invariant) because $\As$ and $\Ad$ are representation of the underlying market states themselves. Figure \ref{fig:global_dual_view} shows a sample of rideshare markets plotted by their respective two-sided views of market equilibrium (with $\Ad$ as the x-axis and $\As$ as the y-axis).

\begin{table}[ht]
    \centering
    \begin{tabular}{|p{0.1\linewidth}||p{0.25\linewidth}|p{0.25\linewidth}|p{0.25\linewidth}|}
    \hline
     $\As \;|\; \Ad$ & Negative & Neutral & Positive \\
     \hline\hline
     Positive & SD misaligned & Over-supplied & Globally over-supplied \\
     \hline
     Neutral & Under-supplied & Globally balanced & Marginally over-supplied \\
     \hline
     Negative & Globally under-supplied & Marginally under-supplied & N.A. \\
     \hline
    \end{tabular}
    \caption{SD balance state classification. The `N.A.' cell follows the results in Proposition \ref{prop:AdLessAs}.}
    \label{tab:state_classification}
\end{table}

We note that the (positive $\Ad$, negative $\As$) cell in Table \ref{tab:state_classification} is marked `N.A.'. In other words, the fourth quadrant of the $(\Ad,\As)$ space is undefined. Here, we explain why.
\begin{proposition}\label{prop:AdLessAs}
As in Section \ref{sec:supply_view}, define $M=\sum_{\ict}\mutildeit$, and $N=\sum_{\ict}\nu_{\ict}$. Given that $M>0,N>0$, the two-sided views of market equilibrium $\Ad$ and $\As$ satisfy $\Ad < \As\cdot\frac{M}{N}$.
\end{proposition}
The proof is provided in Appendix \ref{proof:AdLessAs}.

From Proposition \ref{prop:AdLessAs}, we know that if $\As<0$, then $\Ad<0$. Hence, it is not possible to have $\As<0, \Ad>0$ simultaneously, and no point of $(\Ad,\As)$ can fall into the fourth quadrant. We also note that although it is theoretically possible for a point to be in the (positive $\Ad$, neutral $\As$) or (neutral $\Ad$, negative $\As$) cell in Table \ref{tab:state_classification}, the result in Proposition \ref{prop:AdLessAs} significantly constrains how positive (resp. negative) $\Ad$ (resp. $\As$) can be.

Proposition \ref{prop:AdLessAs} suggests that when $M=N$, $\Ad<\As$. Since $M$ and $N$ being close to each other is a necessary condition for a globally balanced rideshare market, we see that our empirical definition of the neutral intervals in Table \ref{tab:state_classification} is indeed reasonable.

\subsection{A Queuing Perspective}\label{sec:queuing}
The two-sided-view indices $\Ad$ and $\As$ are built on (gh6, 5-min)-level $\mit$, which is the difference between the (unique) open requests and idle driver (allocated to this cell) counts. To understand what the two indices are exactly measuring, it is easier to look at a particular cell from a queuing perspective.\footnote{\cite{zhong2020queueing} show that under assumptions, the system can be analyzed distributedly by cell-level queues.} At any given time, the request in-flow consists of new requests and existing requests from the previous time window not being dropped yet. The request out-flow consists of matched requests and canceled requests. The driver in-flow consists of new drivers sign-in, the idle drivers from the previous time window, drivers who have just finished an order in this cell. The corresponding out-flow consists of matched drivers and drivers sign-off. We assume that requests and drivers are matched one-to-one, and request cancellation and driver sign-off happen before the matching. Any positive difference between the in-flow and out-flow will result in requests and/or drivers stay in the queue to the next time window. 

Since matching is a one-to-one clearance mechanism, a balanced queue for a given cell means that there is little accumulation of demand or supply units from one dispatch cycle to the next. $\mit$ represents a snapshot of this expected accumulation at the end of each 5-min interval. Note that for a given cell, there can only be accumulation of either supply or demand units under the assumption of this queuing framework, whereas at the system level, there can be both due to accumulation of either type of units at the different cells. The GEM indices introduced above are weighted aggregation of the cell-level accumulation conditions. Each $\mit$ measures the difference in the unit counts that have appeared in the two queues over 5-min timescale. 

When considering the region-level system with multiple queues corresponding to the cells, there is additional complexity due to endogeneity of the SD distributions - the driver arrivals at a given time partially depend on the destinations of the requests from earlier time periods. So the demand patterns have an impact on the equilibrium of the queues. It is still an active research question for how endogeneity affects the level of disparity in the two-sided views if the system is left to evolve organically (and hence, how much intervention is required on supply distribution).

\begin{figure}
    \centering
    \includegraphics[width=0.9\textwidth]{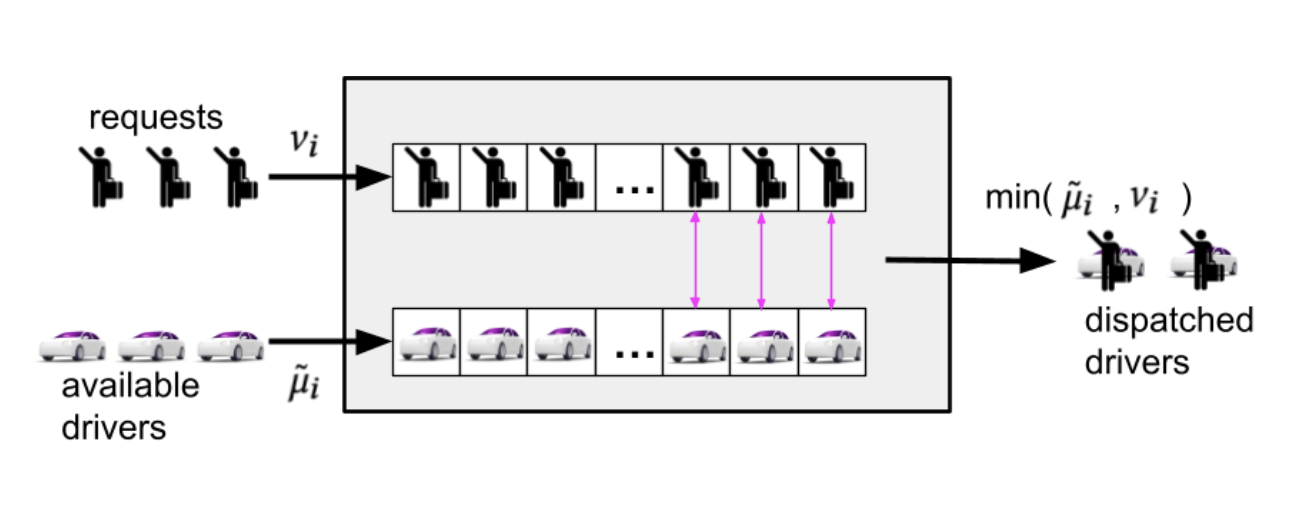}
    \caption{Illustration of a two-sided queue for a given spatiotemporal cell.}
    \label{fig:queuing}
\end{figure}

\section{Market Efficiency}\label{sec:efficiency}
SD-GEM indices $(\Ad, \As)$ can provide a vehicle to measure a market’s efficiency of utilizing supplies for demand fulfillment. Intuitively, it states that in order to sustain a demand-centric balance of $\Ad$ (which drives rider-side metrics and service level), this market has to have a supply-centric balance of $\As$ (which drives supply utilization and earning rate). The origin $(0, 0)$ is the perfectly efficient state because any deviation from it on either the demand or the supply side represents inefficiency (of different degrees) in the market - drivers may be idling more or riders have to wait longer for their requests to be matched. Hence, a marketplace with its $(\Ad, \As)$ `closer' to $(0, 0)$ than another is deemed more efficient. In this section, we investigate the driver and implication of the movement of a point in the 2D space of $(\Ad, \As)$. In Section \ref{sec:estimation}, we go on to show how to quantify the closeness to the origin, and how to test for such changes in experiments.

\subsection{Market State Shift}\label{sec:market_state_shift}
\begin{figure}
    \centering
    \includegraphics[width=0.7\textwidth]{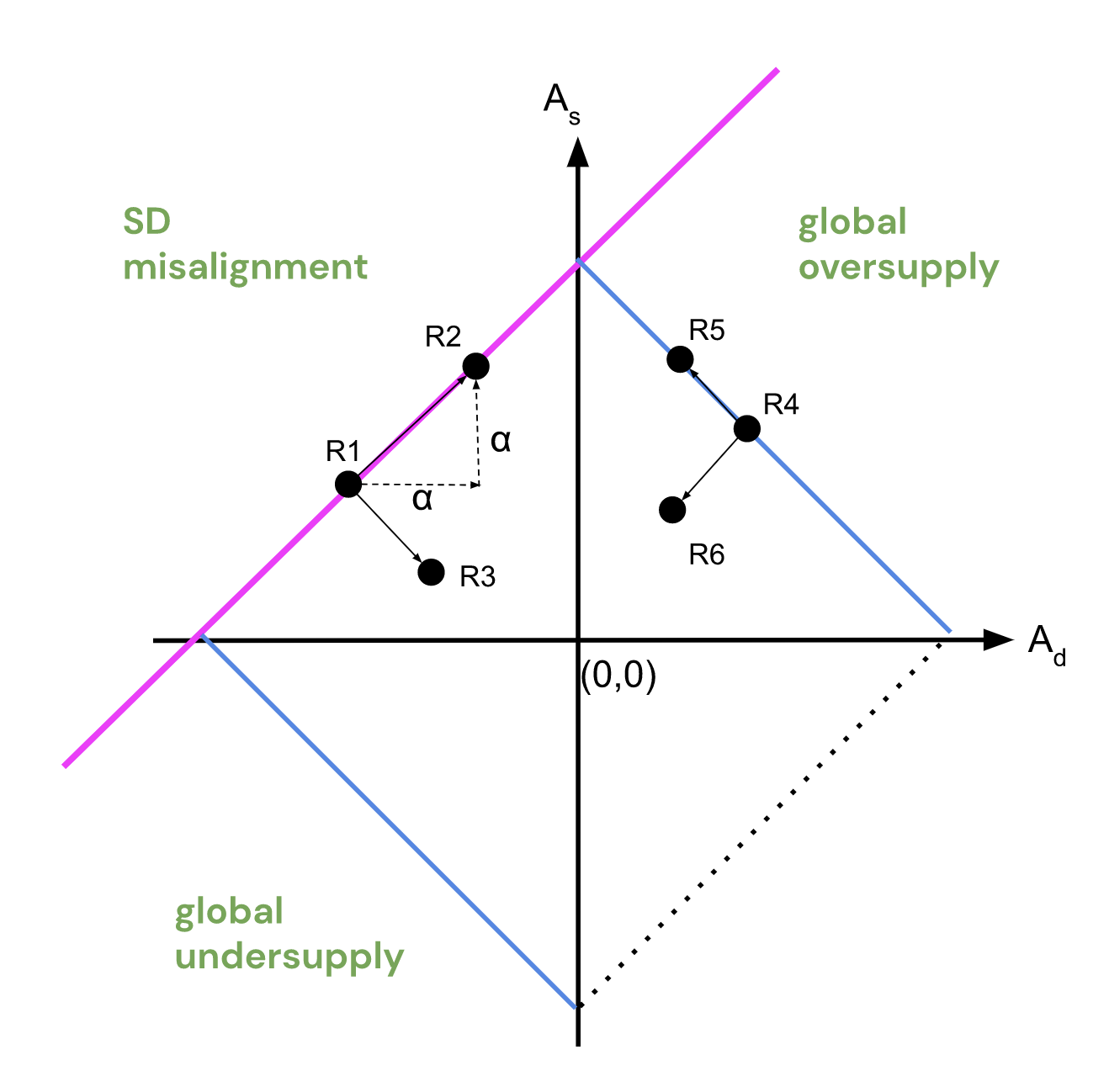}
    \caption{Illustration of different types of market state shifts. The pink line represents the unit-gradient hyperplane that a pure SD volume shift will move the state along. The blue lines are the orthogonal complements representing SD distributional shifts conditional on the market volume.}
    \label{fig:global_frontier}
\end{figure}

To understand movements in the space of the SD-GEM indices, let us examine a specific case of supply influx. Assume that the additional drivers appear in the market in such a way that the dispatched supply distribution follows the current distribution of $\mutilde$ (and hence the dispatched supply per cell increases by a constant fraction $\beta$), then both $\Ad$ and $\As$ will increase by a constant $\alpha=\log(1+\beta)$, illustrated in Figure \ref{fig:global_frontier} by $R_1\rightarrow R_2$ along the pink hyperplane.
Mathematically, the new $\Ad$ after supply influx is 
\begin{equation}\label{eq:Ad_inc}
    \Aprmd = \frac{\sum_{\ict}(\mit+\alpha)\nu_{\ict}}{\sum_{\ict}\nu_{\ict}} = \Ad + \alpha,
\end{equation}
where $\mit + \alpha$ replaces $\mit$ following the definition of $\mit$; and the new $\As$ is
\begin{equation}\label{eq:As_inc}
    \Aprms = \frac{\sum_{\ict}(\mit+\alpha)\muprmtildeit}{\sum_{\ict}\muprmtildeit} = \frac{\sum_{\ict}\mit\muprmtildeit}{\muprmtildeit} + \alpha,
\end{equation}
where $\muprmtilde$ is the optimally dispatched supply (through the GEM optimization ) after the supply influx. By assumption, $\muprmtildeit=\mutildeit(1+\beta)$, we have $\Aprms = \As + \alpha$. 
It is easy to verify that in the case of demand influx, if the additional demand acquired follows the current demand distribution with the same parameter $\beta$, then both $\Ad$ and $\As$ decreases by exactly $\alpha$. Therefore, any SD flux that does not change the current SD distributions will lead to a movement along the pink unit-gradient frontier in Figure \ref{fig:global_frontier}. How much the movement deviates from that depends on the changes in the resulting SD distributions. In general, a movement along the pink line represents an identical change in the expected market state perception of a randomly sampled driver or rider (which again is primarily driven by an organic SD volume shift).

In the second quadrant of the $(\Ad,\As)$ space, a movement along the unit-gradient frontier (e.g., $R_1\rightarrow R_2$) means that an improvement in rider-side experience or service level (a less negative $\Ad$) generally comes with an equal trade-off of a degradation in driver-side metrics such as utilization and earning rate (a more positive $\As$) or vice versa.
The other shift direction of interest is the one orthogonal to the unit-gradient  frontier for organic volume shift. Specifically, $R_1\rightarrow R_3$ in the second quadrant of Figure \ref{fig:global_frontier} is along this orthogonal direction, and it represents a strict Pareto improvement in the $(\Ad,\As)$ space.  It usually requires intervention (e.g., positioning incentives so that supplies are distributed more strategically and more aligned with the demand) to achieve that. In Section \ref{sec:decomp}, we will formalize the interpretation of this movement and utilize the two orthogonal directions in a novel statistical test to capture changes in market efficiency.

\subsection{Implications}
Both of the blue hyperplanes in the first and third quadrants in Figure \ref{fig:global_frontier} are also orthogonal to the pink frontier, together with which and the black dotted hyperplane, they form a contour of the $L^1$-norm.
The filling of the matrix in Table \ref{tab:state_classification} determines that the unit-gradient pink frontier cannot pass through the fourth quadrant of the $(\Ad,\As)$ space.
In the second quadrant, the Manhattan distance w.r.t. the origin does not change if a point moves along the pink frontier, while moving in the orthogonal direction is most effective in terms of reducing the Manhattan distance. This fact suggests that while the two-sided views of market balance are conflicting, it is more effective in terms of improving market efficiency to intervene the SD distributions to address the misalignment problem than acquiring additional supply or demand. On the other hand, in the first and third quadrants, moving along the pink frontier leads to the fastest change in the Manhattan distance (e.g., $R_4\rightarrow R_6$), while moving in the orthogonal direction does not make any difference. This also makes intuitive sense because in the globally over-supplied or under-supplied regimes, SD acquisition or control is a more pressing measure to take. Intervening the SD distributions would not be effective since there is no SD misalignment per se.
In general, bringing any market state to the ideal one (the origin) typically requires both SD volume shift and distributional intervention, corresponding to movements along both of the orthogonal directions, which can be easily seen from Figure \ref{fig:global_frontier}.


\begin{figure}
    \centering
    \includegraphics[width=0.9\textwidth]{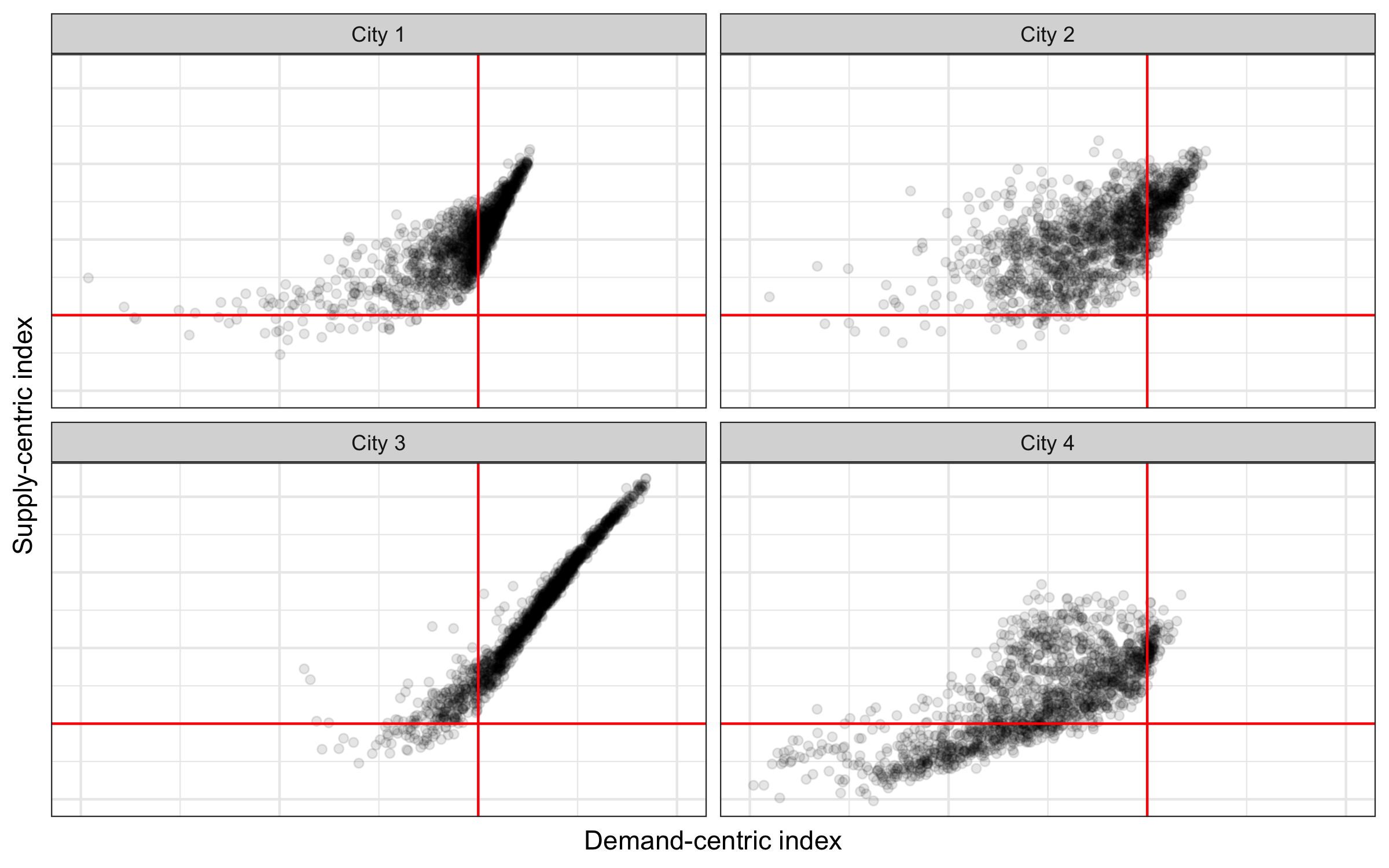}
    \caption{The SD-GEM indices for a sample of four ridesharing markets over a two-month period. Each point represents one hour. The red lines indicate the $x=0$ and $y=0$ axes. We see that the unique topologies and supply and demand patterns of each city give rise to distinct market profiles.}
    \label{fig:city_frontiers}
\end{figure}

\paragraph{Remark}
The OLS regression line in Figure \ref{fig:global_dual_view} resembles the pink line in Figure \ref{fig:global_frontier}, which suggests that across different rideshare markets (of different SD volumes), there are general patterns in the SD spatiotemporal distributions, e.g., city center v.s. suburbs, and morning/evening peak hours.

\paragraph{Remark}
In \cite{zhou2021graph}, an average weighting scheme $\mit := (\nu_{\ict}+\mutildeit)$ for computing the market balance index is also suggested. Intuitively, it is similar to an average of $\Ad$ and $\As$, which also suffers from information loss. In this case, we would not be able to analyze the trade-off between demand and supply-centric views that we have explained in this section and the next.    

\paragraph{Remark}
The SD-GEM framework has two major advantages over existing observational metrics: 1. It has a unified scale of measurement for both supply and demand-side effects, easing the analyses of their trade-offs. 2. It represents the market state in terms of equilibrium and efficiency simultaneously.

\subsection{City-level Profiles}
A closer look at the city-specific plots of the two-sided view in Figure \ref{fig:city_frontiers} reveals that in general, points on the negative side of $\Ad$ have smaller slopes than those on the positive side. An intuitive explanation is that when $\Ad$ is negative, adding more supply into the market would more likely be absorbed by demand (i.e., additional supply is more likely to appear at where supply is relatively low and demand is relatively high, possibly due to the effect of driver incentives). Once $\Ad$ becomes positive, more supply means purely supply surplus (e.g., higher idle rate), and $\As$ increases at a faster rate (which means that additional supply tends to be at where supply is already concentrated). Similar observations apply to either side of $\As$ as well though less obvious. How flat (or steep) the frontier slopes can be depends on the topology of the city and effectiveness of demand/supply shaping strategies because from \eqref{eq:Ad_inc} and \eqref{eq:As_inc}, the spatiotemporal distributions of the incremental (or decremental) supply (or demand) have major impact on the change in the two-sided-view indices. 

City road-network topology has impact on the shape of the points coverage, evident from Figure \ref{fig:city_frontiers}. City 2 is elongated in geographical shape with a small number of arterial roads. Hence, the distribution of supply needs to be strategic to match well with the demand distribution. As a result, there tends to be more variability in $\As$ for any given value of $\Ad$ due to the stochastic nature in the distributions of any incremental supply or demand units, and hence, a less structured $(\Ad,\As)$ profile. City 3 is an example of a dense urban area where streets have high connectivity. This leads to high accessibility among supply and demand units, and there is significantly less variability in $\As$ for any given level of $\Ad$. In other words, the driver and rider's perceptions of market balance are quite predictable from each other.

\section{Estimating efficiency changes}\label{sec:estimation}

Ridesharing platforms constantly iterate on their policies in order to make more efficient use of existing resources. In this section we describe how the discussion from Section~\ref{sec:efficiency} can be leveraged in A/B tests to evaluate different types of efficiency improvement between two candidate policies. In Section~\ref{sec:estimation-manhattan}, we discuss statistics for changes in distance to the origin as a measure of efficiency. In Section~\ref{sec:decomp}, we discuss how to decompose the shift in market state between policies according to the geometry described in Section~\ref{sec:market_state_shift}.

Suppose we have a policy change (treatment) that we wish to compare to the status quo (control) policy. We assume that $\alpha_{d,T}$ and $\alpha_{s,T}$ are the demand- and supply- SD-GEM indices for the treatment condition, and that $\alpha_{d,C}$ and $\alpha_{s,C}$ are the demand- and supply- SD-GEM indices for the control condition. These are considered to be unknown population parameters.

\subsection{Efficiency Improvement}\label{sec:estimation-manhattan}
As described in Section~\ref{sec:efficiency}, the origin (0, 0) represents the perfectly efficient state.  As such, we can consider the estimands
\begin{align}
\tau_1 &= (|\alpha_{d,T}| + |\alpha_{s,T}|) - (|\alpha_{d,C}| + |\alpha_{s,C}|)\label{eqn:tau-manhattan} \\
\tau_2 &= (\alpha_{d,T}^2 + \alpha_{s, T}^2)^{1/2} - (\alpha_{d, C}^2 + \alpha_{s,C}^2)^{1/2} \label{eqn:tau-l2}
\end{align}
These treatment effects measure how much closer to the origin the two-sided-view market state becomes as a result of a policy change. Equation~\eqref{eqn:tau-manhattan} uses the $L^1$ (Manhattan) norm and Equation~\eqref{eqn:tau-l2} uses the $L^2$ (Euclidean) norm. Either norm can be used, but the Manhattan distance ties in directly to the discussion from Section~\ref{sec:market_state_shift} because it induces sparsity with respect to the pink and blue lines in Figure~\ref{fig:global_frontier}. In the second quadrant of the $(\Ad,\As)$ space, $\tau_1 = 0$ for any movement along the pink line (such as $R_1\rightarrow R_2$); in the first and third quadrants, $\tau_1 = 0$ for any movement along the blue line (such as $R_4 \rightarrow R_5$). Both $\tau_1$ and $\tau_2$ are signed quantities such that negative values are desirable and represent improved market health. In what follows and in Section~\ref{sec:applications} we focus on $\tau_1$ only.

\begin{figure}
    \centering
    \includegraphics[width=0.49\linewidth]{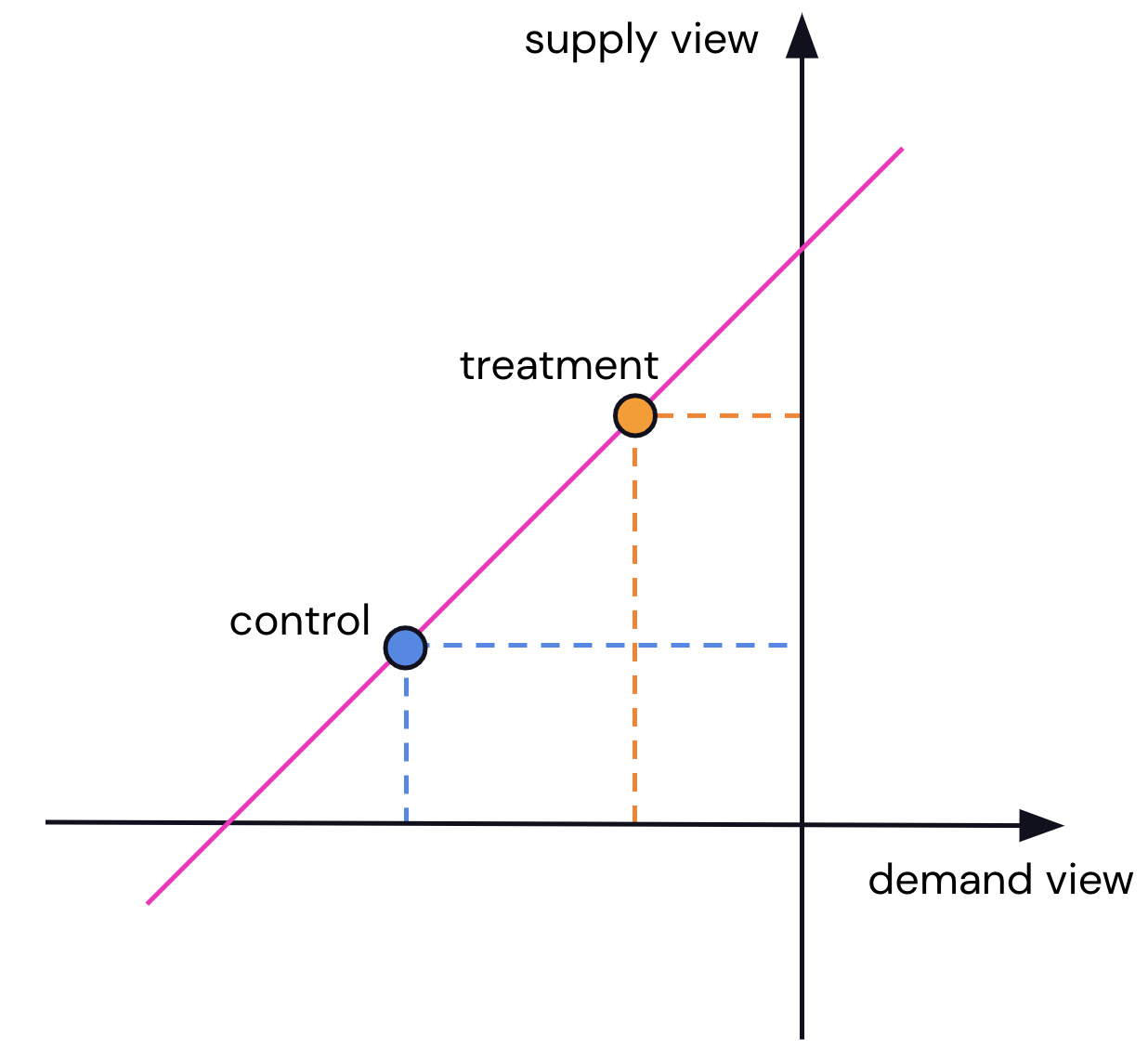}
    \includegraphics[width=0.49\linewidth]{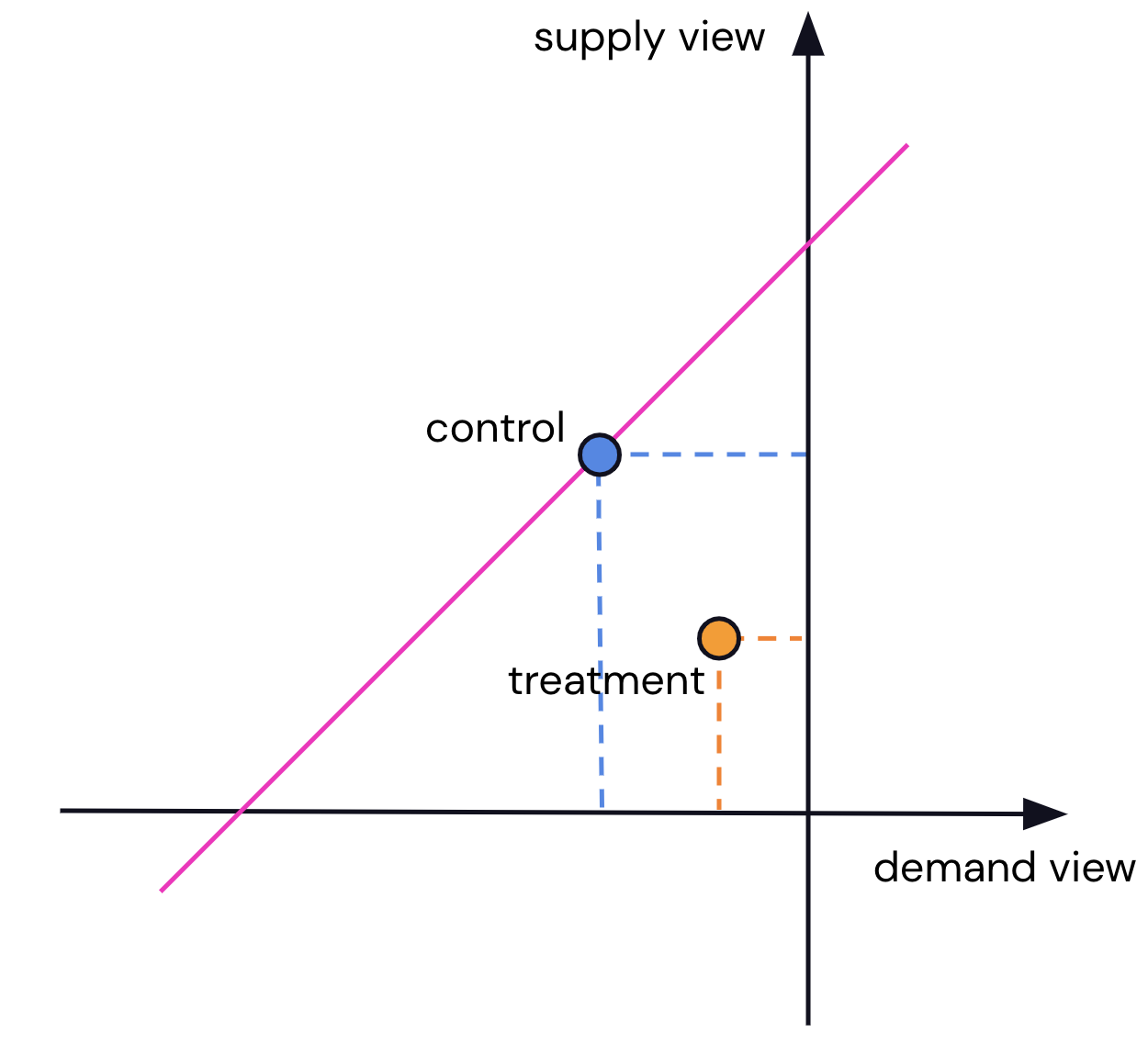}
    \caption{Illustration of $L^1$ treatment effect $\tau_1$. (left) Movement along 45-degree line corresponds to $\tau_1 = 0$ and no improvement in efficiency. (right) Movement closer to origin corresponds to $\tau_1 < 0$.}
    \label{fig:decomp-L1}
\end{figure}

Now, suppose we run an experiment and observe $n_C$ data points from control and $n_T$ data points from treatment, giving us corresponding observations
\begin{align*}
A_{d, C, 1}, \cdots, A_{d, C, n_C} & \qquad A_{s, C, 1}, \cdots, A_{s, C, n_C} \\
A_{d, T, 1}, \cdots, A_{d, T, n_T} & \qquad A_{s, T, 1}, \cdots, A_{s, T, n_T}.
\end{align*}

Then we obtain an estimator for $\tau_1$ by replacing the parameters with their equivalent sample means:
\begin{equation}
\hat \tau_1 = \frac{1}{n_T} \sum_{i=1}^{n_T} (|A_{d, T, i}| + |A_{s, T, i}|) - \frac{1}{n_C} \sum_{i=1}^{n_C} (|A_{d, C, i}| + |A_{s, C, i}|)
\label{eqn:manhattan-estimator}
\end{equation}

\subsection{Decomposing Market State Shift}\label{sec:decomp}

While $\tau_1$ and $\tau_2$ provide the magnitude of efficiency improvement, it is helpful to understand the characteristics of the treatment effects in more detail.  In fact, we can take advantage of the geometry illustrated in Figure~\ref{fig:global_frontier} to measure the relative contributions of volume shift and distributional shift on efficiency gain. Toward that end, consider $\tautotal$ defined by
\[\tautotal^2 = (\alpha_{d,T} - \alpha_{d,C})^2 + (\alpha_{s,T} - \alpha_{s,C})^2.\]
$\tautotal$ is the Euclidean distance between the points $(\alpha_{d,T}, \alpha_{s,T})$ and $(\alpha_{d,C}, \alpha_{s,C})$. (Unlike $\tau_1$ and $\tau_2$, it is always non-negative and thus provides no sense of directionality.) $\tautotal$ captures the magnitude of overall movement from the control policy to the treatment policy.

As discussed in Section~\ref{sec:efficiency}, the unit-gradient line passing through $(\alpha_{d,C}, \alpha_{s,C})$ represents movement primarily due to global supply and demand flux, in which the relative proportion of supply and demand in each spatiotemporal cell does not change from control to treatment. Then we obtain the following proposition based on the projection onto this line.

\begin{figure}
    \centering
    \includegraphics[width=0.49\linewidth]{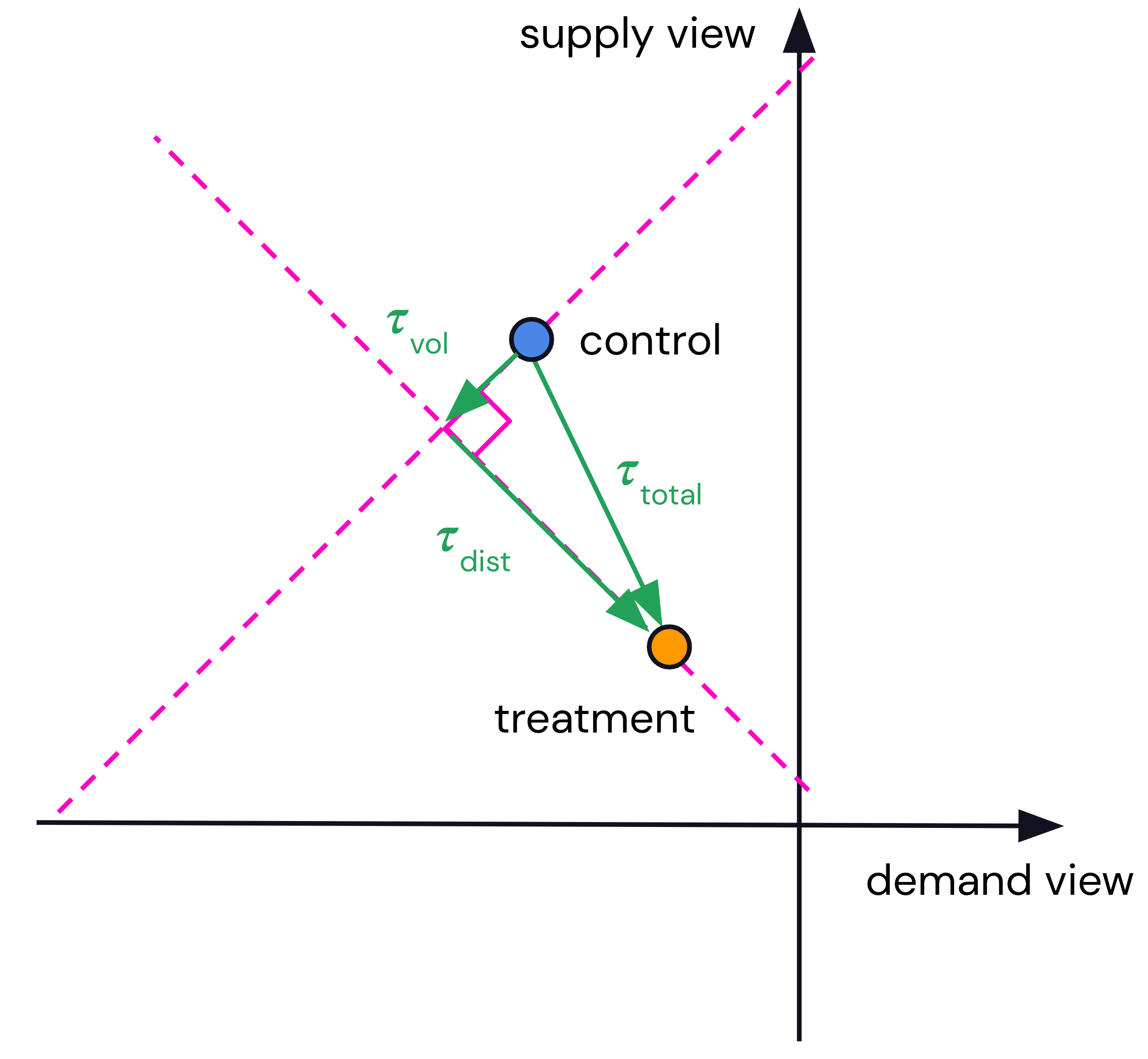}
    \includegraphics[width=0.49\linewidth]{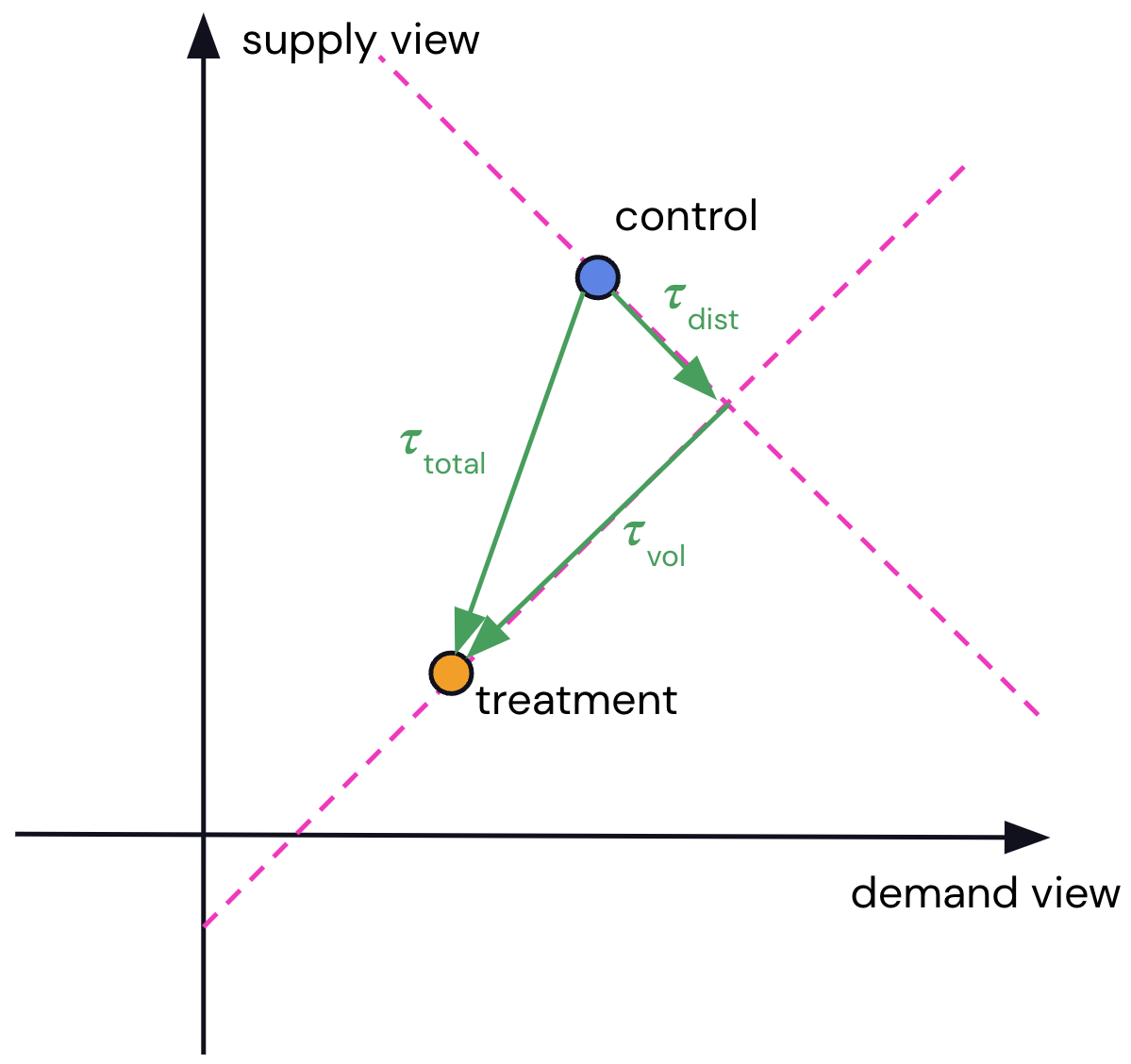}
    \caption{Illustration of decomposition of $\tautotal$. Both diagrams show policy improvements that move $(\Ad, \As)$ closer to the origin. (left) In the second quadrant (SD misalignment), the greatest efficiency improvement comes from more efficient use of resources conditional on existing supply or demand distributions, corresponding to $|\taucond| > |\tauglobal|$. (right) In the first quadrant (global oversupply), the greatest efficiency improvement comes from global demand acquisition, corresponding to $|\tauglobal| > |\taucond|$. The third quadrant (global undersupply) is similar.}
    \label{fig:decomp-L2}
\end{figure}

\begin{proposition}
Define
\begin{align*}
\tauglobal = \frac{1}{\sqrt{2}} \left[ \alpha_{s,T} + \alpha_{d,T} - \alpha_{s,C} - \alpha_{d,C}\right] \\
\taucond = \frac{1}{\sqrt{2}} \left[ \alpha_{s,T} - \alpha_{s,C} - \alpha_{d,T} + \alpha_{d,C}\right].
\end{align*}
Then $\tauglobal$ is the coefficient of the projection of $\tau_2$ onto the unit vector of the 45-degree line with slope +1, and $\taucond$ is the orthogonal complement (the coefficient of the projection onto the unit vector of the line with slope -1). Furthermore,
\[\tautotal^2 = \tauglobal^2 + \taucond^2.\]
\label{prop:decomp}
\end{proposition}
The proof is provided in Appendix \ref{proof:decomp}. $\tauglobal$ corresponds to global volume changes in the demand or supply distributions without modifying the shape of the distribution itself, and $\taucond$ represents efficiency change \emph{conditional} on the existing distribution profile. This decomposition is illustrated in Figure~\ref{fig:decomp-L2}.

As in Equation~\eqref{eqn:manhattan-estimator}, we obtain estimators for these quantities by substituting in the sample means; denoting
\begin{align*}
\bar A_{d, C} = \frac{1}{n_C} \sum_{i=1}^{n_C} A_{d, C, i} & \qquad \bar A_{s, C} = \frac{1}{n_C} \sum_{i=1}^{n_C} A_{s, C, i} \\
\bar A_{d, T} = \frac{1}{n_T} \sum_{i=1}^{n_T} A_{d, T, i} & \qquad \bar A_{s, T} = \frac{1}{n_T} \sum_{i=1}^{n_T} A_{s, T, i}
\end{align*}
we obtain
\begin{align}
\hattautotal &= \sqrt{\left(\bar A_{d, T} - \bar A_{d, C}\right)^2 + \left(\bar A_{s, T} - \bar A_{s, C}\right)^2} \label{eqn:l2-estimator}\\
\hattauglobal &=  \frac{1}{\sqrt{2}} \left[ \bar A_{s,T} + \bar A_{d,T} - \bar A_{s,C} - \bar A_{d,C}\right] \label{eqn:global-estimator}\\
\hattaucond &= \frac{1}{\sqrt{2}} \left[ \bar A_{s,T} - \bar A_{s,C} - \bar A_{d,T} + \bar A_{d,C}\right]. \label{eqn:conditional-estimator}
\end{align}

\subsection{Interpretation}
Notice that in quadrant 1, we have $\tau_1 = \sqrt{2} \tauglobal$, since the demand and supply indices are both positive. Similarly, $\tau_1 = \sqrt{2}\taucond$ in quadrant 2, and $\tau_1 = -\sqrt{2}\tauglobal$ in quadrant 3. These relationships are summarized in Table~\ref{tab:interpretation}. (The interpretation is more complicated when the two policies lie in different quadrants.)
$\tau_1$ takes an all-or-nothing approach to efficiency, which may be more or less desirable depending on the policy being tested and the desired outcomes. In quadrant 2, using $\tau_1$ naturally absorbs any potential confounding from exogenous fluctuations and seasonality in marketplace conditions, which often take the form of changes in the global supply or demand volume. However, it may also be desirable to monitor whether policy changes lead to unintended volume changes.

\begin{table}[ht]
    \centering
    \begin{tabular}{c|c|c|c|c}
    Quadrant & Eff.\ gain/loss & $\tau_1$ & $\tauglobal$ & $\taucond$ \\
    \hline
    \multirow{2}{1cm}{1} & gain & $-$ & $-$ & either \\
    & loss & $+$ & $+$ & either\\
    \hline
    \multirow{2}{1cm}{2} & gain & $-$ & either & $-$ \\
    & loss & $+$ & either & $+$\\
    \hline
    \multirow{2}{1cm}{3} & gain & $-$ & $+$ & either \\
    & loss & $+$ & $-$ & either
    \end{tabular}
    \caption{Relationships among $\tau_1$, $\tauglobal$, and $\taucond$.}
    \label{tab:interpretation}
\end{table}

\section{Empirical results}\label{sec:applications} 
In this section, we apply the estimators defined in Section~\ref{sec:estimation} to perform policy evaluation in A/B tests at Lyft.  We evaluate the policies using switchback experiments~\citep{robins1986new,bojinov2019time,bojinov2022design}. Switchback tests, in which markets are alternatively assigned to treatment and control for a fixed period of time, have become a standard way of conducting A/B tests on ridesharing platforms and other marketplace settings where network effects are prevalent~\citep{chamandy2016experimentation,kastelman2018switchback}.

To use SD-GEM for experiment outcome metrics, we solve the optimization problem~\eqref{eqn:objective} in 5-minute intervals, producing local SD gaps across the duration of the experiment. The demand- and supply-centric views are then calculated for each time period by aggregating the local SD gaps temporally across the duration of the time period and spatially across the geographical extent of the market. The estimators defined in Equations~\eqref{eqn:manhattan-estimator}, \eqref{eqn:l2-estimator}, \eqref{eqn:global-estimator}, and \eqref{eqn:conditional-estimator} can be calculated using these observations.

Statistical inference can be performed using permutation tests, in which treatment and control groups are repeatedly re-sampled from the distribution of possible experimental designs. The test statistic is computed for each permutation sample, producing a null distribution. The $p$-value is taken to be the proportion of the null distribution more extreme than the observed statistic. The empirical standard deviation of the permutation distribution is taken as the standard error for the estimator.

We examine results for two experiments, both of which were run for two weeks. Experiment 1, run in 14 markets, is a change to the dynamic pricing algorithm, with a period length of 120 minutes and thus 168 observations per market over the two-week experiment, so that $n_C = n_T = 1176$. Experiment 2, run in 4 markets, involves a joint change to the driver pay and rider pricing policies and has periods of 56 minutes, yielding 360 observations per market over the two-week experiment, so that $n_C = n_T = 720$. For $\tautotal$ we compute a one-sided $p$-value since $\tautotal$ is always non-negative; for all others we compute two-sided $p$-values. We use 1000 permutation samples.  The results are shown in Table~\ref{tab:expt_estimates}. Also included is the optimal GEM objective value \eqref{eqn:objective}, denoted by $\rho$, which \cite{zhou2021graph} suggest using as a test statistic. 

The experiments were run in markets where the status quo was in Quadrant 2 from Table~\ref{tab:interpretation} (SD misalignment). Although the point estimate for $\rho$ is negative, corresponding to improved efficiency, it captures only SD coherence and has low power, especially given that these are two substantial policy changes with large effect sizes in key operational metrics. Nor does $\rho$ provide much insight into the source of efficiency improvement. For Experiment 1, $A_d$ is statistically significant, whereas $A_s$ is not and has smaller magnitude. This is expected because Experiment 1 involves a change in pricing policy, which directly affects riders and only indirectly affects drivers. The estimates for $\tau_2$ show that the improvement for Experiment 2 is driven primarily by a shift that does not change the existing supply and demand distributions, whereas in Experiment 1 $\tauglobal$ and $\taucond$ are both statistically significant. As such, Experiment 2 functions primarily to increase overall supply and decrease overall demand, and indeed, treatment effects for operational metrics in Experiment 2 show that rider-side outcomes (ETAs and cancellation rates) are improved where as driver-side outcomes (such as idling time) are degraded. In contrast, Experiment 1 is an improvement to the dynamic pricing algorithm and thus directly shapes the request distribution. Correspondingly, we see an improvement in rider-side outcomes whereas driver-side outcomes are not statistically significant.

\begin{figure}
    \centering
    \includegraphics[width=0.8\linewidth]{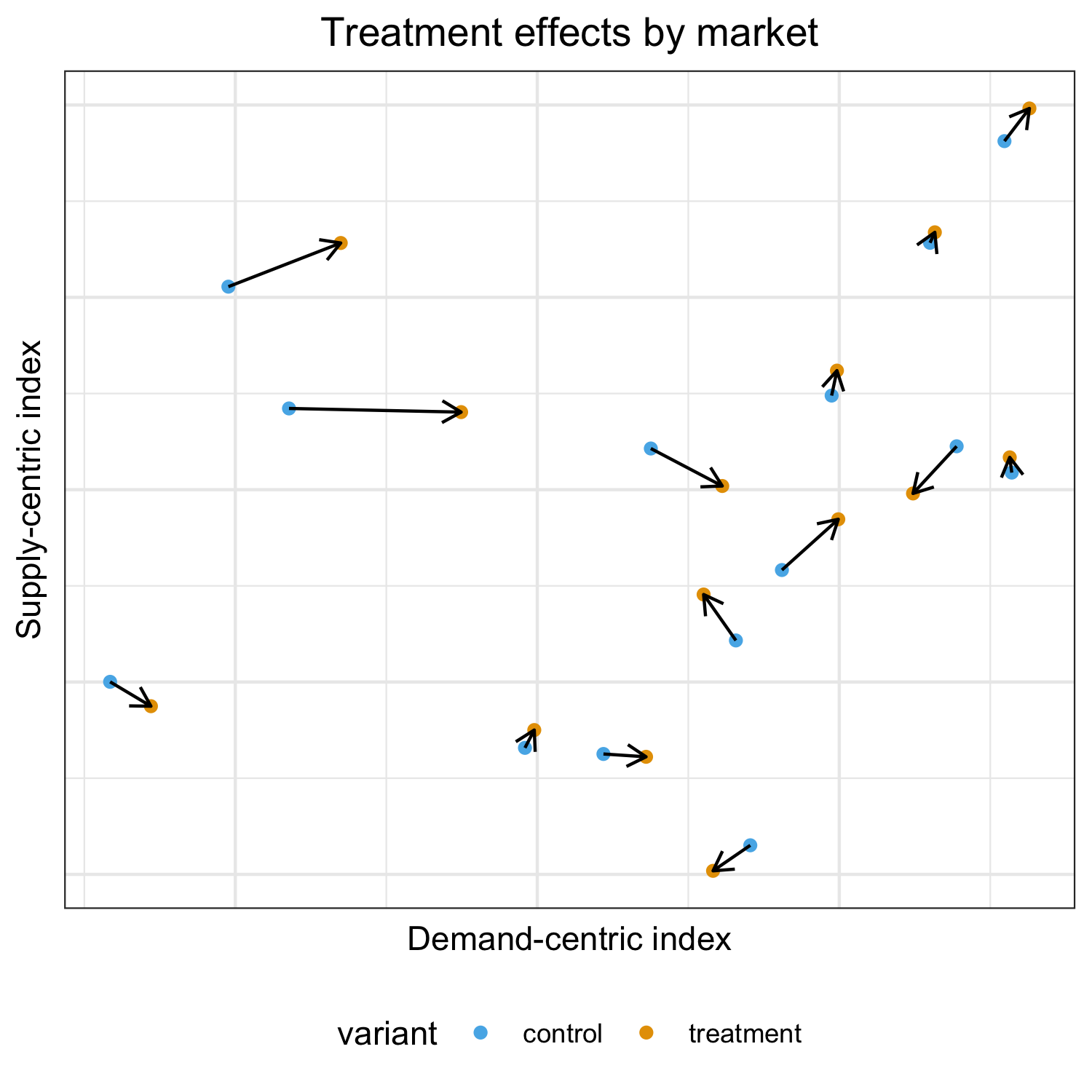}
    \caption{Observed movement in SD-GEM indices for Experiment 1, separated by market.}
    \label{fig:ate_by_market}
\end{figure}

In Figure~\ref{fig:ate_by_market}, we plot the movement of the $(\Ad, \As)$ vectors for Experiment 1 separated by market. The aggregate movement is rightward and upward, which corresponds with the treatment effects presented in Table~\ref{tab:expt_estimates}.  However, there is a good amount of heterogeneity by market, suggesting that a more sophisticated policy optimization may be beneficial.

\begin{table}[ht]
    \centering
    \begin{tabular}{c|c|c|c|c|c|c}
    \multirow{2}{1cm}{Statistic} & \multicolumn{3}{c|}{Experiment 1}& \multicolumn{3}{c}{Experiment 2}\\
\cline{2-7}
    & Est. & SE & $p$-value & Est. & SE & $p$-value\\
     \hline
     $\rho$ & -3.592 & 3.295 & 0.280 & -4.941 & 5.276 & 0.545 \\
     \hline
     $A_d$  & 0.022 & 0.009 & \textbf{0.015} & 0.036 & 0.021 & \textbf{0.022}  \\
     \hline
     $A_s$ & 0.004 & 0.004 & 0.228 & 0.014 & 0.006 & \textbf{0.025} \\
     \hline
     $\tau_1$  & -0.018 & 0.008 & \textbf{0.025} & -0.019 & 0.022 & 0.560 \\
     \hline
     $\tautotal$   & 0.022 & 0.005 & \textbf{0.014} & 0.039 & 0.008 & \textbf{0.014} \\
     \hline
     $\tauglobal$   & 0.019 & 0.007 & \textbf{0.013} & 0.035 & 0.016 & \textbf{0.001} \\
     \hline
     $\taucond$   & -0.013 & 0.006 & \textbf{0.039} & -0.016 & 0.014 & 0.317 \\
    \end{tabular}
    \caption{Point estimates, standard errors, and $p$-values for two ridesharing experiments. $p$-values that are significant at the $0.05$ level are highlighted in bold.}
    \label{tab:expt_estimates}
\end{table}


\section{Conclusion}\label{sec:conclusion}

In this work, we build upon GEM proposed by~\citep{zhou2021graph} and develop SD-GEM, a unified representation framework for rideshare marketplace equilibrium and efficiency based on a two-sided view of the market state in ridesharing platforms. We characterize the difference between global volume shifts in supply and demand and targeted distributional shifts, via the geometry of the two-dimensional SD-GEM market indices. We show how this perspective can be used for policy evaluation by developing a novel suite of estimators for A/B tests for measuring efficiency improvement and decomposing market shift into these two components.

There are several avenues that we have not addressed in this work but which are promising for future studies.  First, this work studies the simplest rideshare setting comprised of only standard single-occupancy rides. However, ridesharing platforms offer a multi-modal product in which passengers select from multiple modes (e.g. standard, luxury, or shared rides) and drivers mark themselves as eligible for one or more of these types of rides. Also related is the case of linked rides, in which a queue of rides is assigned to a single driver to minimize idle time.  Second, the queuing perspective discussed in Section~\ref{sec:queuing} can be extended by considering an additional asymmetry between riders and drivers. While matching is a one-to-one clearance mechanism, remaining in the queue for extended periods of time is a much more negative experience for riders than for drivers (riders expect to be matched within a few seconds, whereas drivers can be idle for several minutes). Understanding this asymmetry and its implications for ridesharing can provide further perspective on marketplace efficiency. Finally, one can consider the extent to which the insights here can be applied to other types of two-sided marketplaces. For example, in e-commerce markets that manage relationships between merchants and customers, an understanding of customer-focused and merchant-focused marketplace efficiency and the interaction between them would still be useful for guiding incentive spend and making policy decisions, even though the transportation component that is critical to ridesharing is absent.

\section{Acknowledgements}
The authors would like to thank Ricky Chachra, Mark Huberty, John Kirn, Ido Bright, and Nicholas Chamandy at Lyft for their support and for insightful discussions on topics relevant to this paper.  
\bibliographystyle{ACM-Reference-Format}
\bibliography{gem}


\begin{thebibliography}{34}


\ifx \showCODEN    \undefined \def \showCODEN     #1{\unskip}     \fi
\ifx \showDOI      \undefined \def \showDOI       #1{#1}\fi
\ifx \showISBNx    \undefined \def \showISBNx     #1{\unskip}     \fi
\ifx \showISBNxiii \undefined \def \showISBNxiii  #1{\unskip}     \fi
\ifx \showISSN     \undefined \def \showISSN      #1{\unskip}     \fi
\ifx \showLCCN     \undefined \def \showLCCN      #1{\unskip}     \fi
\ifx \shownote     \undefined \def \shownote      #1{#1}          \fi
\ifx \showarticletitle \undefined \def \showarticletitle #1{#1}   \fi
\ifx \showURL      \undefined \def \showURL       {\relax}        \fi
\providecommand\bibfield[2]{#2}
\providecommand\bibinfo[2]{#2}
\providecommand\natexlab[1]{#1}
\providecommand\showeprint[2][]{arXiv:#2}

\bibitem[Al-Chanati and Iyer(2021)]%
        {al2021drives}
\bibfield{author}{\bibinfo{person}{Motaz Al-Chanati} {and}
  \bibinfo{person}{Vinayak Iyer}.} \bibinfo{year}{2021}\natexlab{}.
\newblock \showarticletitle{What Drives the Efficiency in Ridesharing Markets?
  Evidence from Austin, Texas}.
\newblock \bibinfo{journal}{\emph{Evidence from Austin, Texas (November 8,
  2021)}} (\bibinfo{year}{2021}).
\newblock


\bibitem[Bimpikis et~al\mbox{.}(2019)]%
        {bimpikis2019spatial}
\bibfield{author}{\bibinfo{person}{Kostas Bimpikis}, \bibinfo{person}{Ozan
  Candogan}, {and} \bibinfo{person}{Daniela Saban}.}
  \bibinfo{year}{2019}\natexlab{}.
\newblock \showarticletitle{Spatial pricing in ride-sharing networks}.
\newblock \bibinfo{journal}{\emph{Operations Research}} \bibinfo{volume}{67},
  \bibinfo{number}{3} (\bibinfo{year}{2019}), \bibinfo{pages}{744--769}.
\newblock


\bibitem[Bojinov and Shephard(2019)]%
        {bojinov2019time}
\bibfield{author}{\bibinfo{person}{Iavor Bojinov} {and} \bibinfo{person}{Neil
  Shephard}.} \bibinfo{year}{2019}\natexlab{}.
\newblock \showarticletitle{Time series experiments and causal estimands: exact
  randomization tests and trading}.
\newblock \bibinfo{journal}{\emph{J. Amer. Statist. Assoc.}}
  \bibinfo{volume}{114}, \bibinfo{number}{528} (\bibinfo{year}{2019}),
  \bibinfo{pages}{1665--1682}.
\newblock


\bibitem[Bojinov et~al\mbox{.}(2022)]%
        {bojinov2022design}
\bibfield{author}{\bibinfo{person}{Iavor Bojinov}, \bibinfo{person}{David
  Simchi-Levi}, {and} \bibinfo{person}{Jinglong Zhao}.}
  \bibinfo{year}{2022}\natexlab{}.
\newblock \showarticletitle{Design and analysis of switchback experiments}.
\newblock \bibinfo{journal}{\emph{Management Science}} (\bibinfo{year}{2022}).
\newblock


\bibitem[Braverman et~al\mbox{.}(2019)]%
        {braverman2019empty}
\bibfield{author}{\bibinfo{person}{Anton Braverman}, \bibinfo{person}{Jim~G
  Dai}, \bibinfo{person}{Xin Liu}, {and} \bibinfo{person}{Lei Ying}.}
  \bibinfo{year}{2019}\natexlab{}.
\newblock \showarticletitle{Empty-car routing in ridesharing systems}.
\newblock \bibinfo{journal}{\emph{Operations Research}} \bibinfo{volume}{67},
  \bibinfo{number}{5} (\bibinfo{year}{2019}), \bibinfo{pages}{1437--1452}.
\newblock


\bibitem[Chamandy(2016)]%
        {chamandy2016experimentation}
\bibfield{author}{\bibinfo{person}{Nicholas Chamandy}.}
  \bibinfo{year}{2016}\natexlab{}.
\newblock \bibinfo{title}{Experimentation in a Ridesharing Marketplace}.
\newblock
  \bibinfo{howpublished}{\url{https://eng.lyft.com/experimentation-in-a-ridesharing-marketplace-f75a9c4fcf01}}.
\newblock
\newblock
\shownote{Accessed: 2023-01-18}.


\bibitem[Chaudhari et~al\mbox{.}(2020)]%
        {chaudhari2020learn}
\bibfield{author}{\bibinfo{person}{Harshal~A Chaudhari},
  \bibinfo{person}{John~W Byers}, {and} \bibinfo{person}{Evimaria Terzi}.}
  \bibinfo{year}{2020}\natexlab{}.
\newblock \showarticletitle{Learn to Earn: Enabling Coordination within a Ride
  Hailing Fleet}.
\newblock \bibinfo{journal}{\emph{Proceedings of IEEE International Conference
  on Big Data}} (\bibinfo{year}{2020}).
\newblock


\bibitem[Chen et~al\mbox{.}(2021)]%
        {chen2021spatial}
\bibfield{author}{\bibinfo{person}{Chuqiao Chen}, \bibinfo{person}{Fugen Yao},
  \bibinfo{person}{Dong Mo}, \bibinfo{person}{Jiangtao Zhu}, {and}
  \bibinfo{person}{Xiqun~Michael Chen}.} \bibinfo{year}{2021}\natexlab{}.
\newblock \showarticletitle{Spatial-temporal pricing for ride-sourcing platform
  with reinforcement learning}.
\newblock \bibinfo{journal}{\emph{Transportation Research Part C: Emerging
  Technologies}}  \bibinfo{volume}{130} (\bibinfo{year}{2021}),
  \bibinfo{pages}{103272}.
\newblock


\bibitem[Eshkevari et~al\mbox{.}(2022)]%
        {eshkevari2022reinforcement}
\bibfield{author}{\bibinfo{person}{Soheil~Sadeghi Eshkevari},
  \bibinfo{person}{Xiaocheng Tang}, \bibinfo{person}{Zhiwei Qin},
  \bibinfo{person}{Jinhan Mei}, \bibinfo{person}{Cheng Zhang},
  \bibinfo{person}{Qianying Meng}, {and} \bibinfo{person}{Jia Xu}.}
  \bibinfo{year}{2022}\natexlab{}.
\newblock \showarticletitle{Reinforcement Learning in the Wild: Scalable RL
  Dispatching Algorithm Deployed in Ridehailing Marketplace}. In
  \bibinfo{booktitle}{\emph{Proceedings of the 26th ACM SIGKDD International
  Conference on Knowledge Discovery \& Data Mining}}.
\newblock


\bibitem[Geng et~al\mbox{.}(2019)]%
        {geng2019spatiotemporal}
\bibfield{author}{\bibinfo{person}{Xu Geng}, \bibinfo{person}{Yaguang Li},
  \bibinfo{person}{Leye Wang}, \bibinfo{person}{Lingyu Zhang},
  \bibinfo{person}{Qiang Yang}, \bibinfo{person}{Jieping Ye}, {and}
  \bibinfo{person}{Yan Liu}.} \bibinfo{year}{2019}\natexlab{}.
\newblock \showarticletitle{Spatiotemporal multi-graph convolution network for
  ride-hailing demand forecasting}. In \bibinfo{booktitle}{\emph{Proceedings of
  the AAAI conference on artificial intelligence}}, Vol.~\bibinfo{volume}{33}.
  \bibinfo{pages}{3656--3663}.
\newblock


\bibitem[Ghili and Kumar(2021)]%
        {ghili2021spatial}
\bibfield{author}{\bibinfo{person}{Soheil Ghili} {and} \bibinfo{person}{Vineet
  Kumar}.} \bibinfo{year}{2021}\natexlab{}.
\newblock \showarticletitle{Spatial distribution of supply and the role of
  market thickness: Theory and evidence from ride sharing}.
\newblock \bibinfo{journal}{\emph{arXiv preprint arXiv:2108.05954}}
  (\bibinfo{year}{2021}).
\newblock


\bibitem[Han et~al\mbox{.}(2022)]%
        {han2022real}
\bibfield{author}{\bibinfo{person}{Benjamin Han}, \bibinfo{person}{Hyungjun
  Lee}, {and} \bibinfo{person}{S{\'e}bastien Martin}.}
  \bibinfo{year}{2022}\natexlab{}.
\newblock \showarticletitle{Real-Time Rideshare Driver Supply Values Using
  Online Reinforcement Learning}. In \bibinfo{booktitle}{\emph{Proceedings of
  the 28th ACM SIGKDD Conference on Knowledge Discovery and Data Mining}}.
  \bibinfo{pages}{2968--2976}.
\newblock


\bibitem[Han et~al\mbox{.}(2017)]%
        {han2017quantification}
\bibfield{author}{\bibinfo{person}{Shuo Han}, \bibinfo{person}{Ufuk Topcu},
  {and} \bibinfo{person}{George~J Pappas}.} \bibinfo{year}{2017}\natexlab{}.
\newblock \showarticletitle{Quantification on the efficiency gain of automated
  ridesharing services}. In \bibinfo{booktitle}{\emph{2017 American Control
  Conference (ACC)}}. IEEE, \bibinfo{pages}{3560--3566}.
\newblock


\bibitem[Hu et~al\mbox{.}(2022)]%
        {hu2021surge}
\bibfield{author}{\bibinfo{person}{Bin Hu}, \bibinfo{person}{Ming Hu}, {and}
  \bibinfo{person}{Han Zhu}.} \bibinfo{year}{2022}\natexlab{}.
\newblock \showarticletitle{Surge pricing and two-sided temporal responses in
  ride hailing}.
\newblock \bibinfo{journal}{\emph{Manufacturing \& Service Operations
  Management}} \bibinfo{volume}{24}, \bibinfo{number}{1}
  (\bibinfo{year}{2022}), \bibinfo{pages}{91--109}.
\newblock


\bibitem[Jiao et~al\mbox{.}(2021)]%
        {jtq2021repos}
\bibfield{author}{\bibinfo{person}{Yan Jiao}, \bibinfo{person}{Xiaocheng Tang},
  \bibinfo{person}{Zhiwei~(Tony) Qin}, \bibinfo{person}{Shuaiji Li},
  \bibinfo{person}{Fan Zhang}, \bibinfo{person}{Hongtu Zhu}, {and}
  \bibinfo{person}{Jieping Ye}.} \bibinfo{year}{2021}\natexlab{}.
\newblock \showarticletitle{Real-world ride-hailing vehicle repositioning using
  deep reinforcement learning}.
\newblock \bibinfo{journal}{\emph{Transportation Research Part C: Emerging
  Technologies}}  \bibinfo{volume}{130} (\bibinfo{year}{2021}),
  \bibinfo{pages}{103289}.
\newblock
\showISSN{0968-090X}
\urldef\tempurl%
\url{https://doi.org/10.1016/j.trc.2021.103289}
\showDOI{\tempurl}


\bibitem[Kastelman and Ramesh(2018)]%
        {kastelman2018switchback}
\bibfield{author}{\bibinfo{person}{David Kastelman} {and}
  \bibinfo{person}{Raghav Ramesh}.} \bibinfo{year}{2018}\natexlab{}.
\newblock \bibinfo{title}{Switchback Tests and Randomized Experimentation Under
  Network Effects at DoorDash}.
\newblock
  \bibinfo{howpublished}{\url{https://medium.com/@DoorDash/switchback-tests-and-randomized-experimentation-under-network-effects-at-doordash-f1d938ab7c2a}}.
\newblock
\newblock
\shownote{Accessed: 2023-01-18}.


\bibitem[Ke et~al\mbox{.}(2020)]%
        {ke2020pricing}
\bibfield{author}{\bibinfo{person}{Jintao Ke}, \bibinfo{person}{Hai Yang},
  \bibinfo{person}{Xinwei Li}, \bibinfo{person}{Hai Wang}, {and}
  \bibinfo{person}{Jieping Ye}.} \bibinfo{year}{2020}\natexlab{}.
\newblock \showarticletitle{Pricing and equilibrium in on-demand ride-pooling
  markets}.
\newblock \bibinfo{journal}{\emph{Transportation Research Part B:
  Methodological}}  \bibinfo{volume}{139} (\bibinfo{year}{2020}),
  \bibinfo{pages}{411--431}.
\newblock


\bibitem[Lin et~al\mbox{.}(2018)]%
        {lin2018efficient}
\bibfield{author}{\bibinfo{person}{Kaixiang Lin}, \bibinfo{person}{Renyu Zhao},
  \bibinfo{person}{Zhe Xu}, {and} \bibinfo{person}{Jiayu Zhou}.}
  \bibinfo{year}{2018}\natexlab{}.
\newblock \showarticletitle{Efficient large-scale fleet management via
  multi-agent deep reinforcement learning}. In
  \bibinfo{booktitle}{\emph{Proceedings of the 24th ACM SIGKDD International
  Conference on Knowledge Discovery \& Data Mining}}.
  \bibinfo{pages}{1774--1783}.
\newblock


\bibitem[Ma et~al\mbox{.}(2022)]%
        {ma2022spatio}
\bibfield{author}{\bibinfo{person}{Hongyao Ma}, \bibinfo{person}{Fei Fang},
  {and} \bibinfo{person}{David~C Parkes}.} \bibinfo{year}{2022}\natexlab{}.
\newblock \showarticletitle{Spatio-temporal pricing for ridesharing platforms}.
\newblock \bibinfo{journal}{\emph{Operations Research}} \bibinfo{volume}{70},
  \bibinfo{number}{2} (\bibinfo{year}{2022}), \bibinfo{pages}{1025--1041}.
\newblock


\bibitem[Ong et~al\mbox{.}(2021)]%
        {ong2021driver}
\bibfield{author}{\bibinfo{person}{Hao~Yi Ong}, \bibinfo{person}{Daniel
  Freund}, {and} \bibinfo{person}{Davide Crapis}.}
  \bibinfo{year}{2021}\natexlab{}.
\newblock \showarticletitle{Driver Positioning and Incentive Budgeting with an
  Escrow Mechanism for Ride-Sharing Platforms}.
\newblock \bibinfo{journal}{\emph{INFORMS Journal on Applied Analytics}}
  \bibinfo{volume}{51}, \bibinfo{number}{5} (\bibinfo{year}{2021}),
  \bibinfo{pages}{373--390}.
\newblock


\bibitem[Qin et~al\mbox{.}(2022)]%
        {qin2022reinforcement}
\bibfield{author}{\bibinfo{person}{Zhiwei~Tony Qin}, \bibinfo{person}{Hongtu
  Zhu}, {and} \bibinfo{person}{Jieping Ye}.} \bibinfo{year}{2022}\natexlab{}.
\newblock \showarticletitle{Reinforcement learning for ridesharing: An extended
  survey}.
\newblock \bibinfo{journal}{\emph{Transportation Research Part C: Emerging
  Technologies}}  \bibinfo{volume}{144} (\bibinfo{year}{2022}),
  \bibinfo{pages}{103852}.
\newblock


\bibitem[Robins(1986)]%
        {robins1986new}
\bibfield{author}{\bibinfo{person}{James Robins}.}
  \bibinfo{year}{1986}\natexlab{}.
\newblock \showarticletitle{A new approach to causal inference in mortality
  studies with a sustained exposure period—application to control of the
  healthy worker survivor effect}.
\newblock \bibinfo{journal}{\emph{Mathematical modelling}} \bibinfo{volume}{7},
  \bibinfo{number}{9-12} (\bibinfo{year}{1986}), \bibinfo{pages}{1393--1512}.
\newblock


\bibitem[Shang et~al\mbox{.}(2019)]%
        {shang2019environment}
\bibfield{author}{\bibinfo{person}{Wenjie Shang}, \bibinfo{person}{Yang Yu},
  \bibinfo{person}{Qingyang Li}, \bibinfo{person}{Zhiwei Qin},
  \bibinfo{person}{Yiping Meng}, {and} \bibinfo{person}{Jieping Ye}.}
  \bibinfo{year}{2019}\natexlab{}.
\newblock \showarticletitle{Environment reconstruction with hidden confounders
  for reinforcement learning based recommendation}. In
  \bibinfo{booktitle}{\emph{Proceedings of the 25th ACM SIGKDD International
  Conference on Knowledge Discovery \& Data Mining}}.
  \bibinfo{pages}{566--576}.
\newblock


\bibitem[Tang et~al\mbox{.}(2019)]%
        {tang2019deep}
\bibfield{author}{\bibinfo{person}{Xiaocheng Tang}, \bibinfo{person}{Zhiwei
  Qin}, \bibinfo{person}{Fan Zhang}, \bibinfo{person}{Zhaodong Wang},
  \bibinfo{person}{Zhe Xu}, \bibinfo{person}{Yintai Ma},
  \bibinfo{person}{Hongtu Zhu}, {and} \bibinfo{person}{Jieping Ye}.}
  \bibinfo{year}{2019}\natexlab{}.
\newblock \showarticletitle{A deep value-network based approach for
  multi-driver order dispatching}. In \bibinfo{booktitle}{\emph{Proceedings of
  the 25th ACM SIGKDD international conference on knowledge discovery \& data
  mining}}. \bibinfo{pages}{1780--1790}.
\newblock


\bibitem[Tang et~al\mbox{.}(2021)]%
        {tang2021value}
\bibfield{author}{\bibinfo{person}{Xiaocheng Tang}, \bibinfo{person}{Fan
  Zhang}, \bibinfo{person}{Zhiwei Qin}, \bibinfo{person}{Yansheng Wang},
  \bibinfo{person}{Dingyuan Shi}, \bibinfo{person}{Bingchen Song},
  \bibinfo{person}{Yongxin Tong}, \bibinfo{person}{Hongtu Zhu}, {and}
  \bibinfo{person}{Jieping Ye}.} \bibinfo{year}{2021}\natexlab{}.
\newblock \showarticletitle{Value Function is All You Need: A Unified Learning
  Framework for Ride Hailing Platforms}. In
  \bibinfo{booktitle}{\emph{Proceedings of the 27th ACM SIGKDD Conference on
  Knowledge Discovery \& Data Mining}} (Virtual Event, Singapore)
  \emph{(\bibinfo{series}{KDD '21})}. \bibinfo{publisher}{Association for
  Computing Machinery}, \bibinfo{address}{New York, NY, USA},
  \bibinfo{pages}{3605–3615}.
\newblock
\showISBNx{9781450383325}
\urldef\tempurl%
\url{https://doi.org/10.1145/3447548.3467096}
\showDOI{\tempurl}


\bibitem[Tong et~al\mbox{.}(2017)]%
        {tong2017simpler}
\bibfield{author}{\bibinfo{person}{Yongxin Tong}, \bibinfo{person}{Yuqiang
  Chen}, \bibinfo{person}{Zimu Zhou}, \bibinfo{person}{Lei Chen},
  \bibinfo{person}{Jie Wang}, \bibinfo{person}{Qiang Yang},
  \bibinfo{person}{Jieping Ye}, {and} \bibinfo{person}{Weifeng Lv}.}
  \bibinfo{year}{2017}\natexlab{}.
\newblock \showarticletitle{The simpler the better: a unified approach to
  predicting original taxi demands based on large-scale online platforms}. In
  \bibinfo{booktitle}{\emph{Proceedings of the 23rd ACM SIGKDD international
  conference on knowledge discovery and data mining}}.
  \bibinfo{pages}{1653--1662}.
\newblock


\bibitem[Wang and Yang(2019)]%
        {wang2019ridesourcing}
\bibfield{author}{\bibinfo{person}{Hai Wang} {and} \bibinfo{person}{Hai Yang}.}
  \bibinfo{year}{2019}\natexlab{}.
\newblock \showarticletitle{Ridesourcing systems: A framework and review}.
\newblock \bibinfo{journal}{\emph{Transportation Research Part B:
  Methodological}}  \bibinfo{volume}{129} (\bibinfo{year}{2019}),
  \bibinfo{pages}{122--155}.
\newblock


\bibitem[Wu et~al\mbox{.}(2022)]%
        {wu2022spatio}
\bibfield{author}{\bibinfo{person}{Yanqiu Wu}, \bibinfo{person}{Qingyang Li},
  {and} \bibinfo{person}{Zhiwei Qin}.} \bibinfo{year}{2022}\natexlab{}.
\newblock \showarticletitle{Spatio-temporal Incentives Optimization for
  Ride-hailing Services with Offline Deep Reinforcement Learning}.
\newblock \bibinfo{journal}{\emph{arXiv preprint arXiv:2211.03240}}
  (\bibinfo{year}{2022}).
\newblock


\bibitem[Xu et~al\mbox{.}(2018)]%
        {xu2018large}
\bibfield{author}{\bibinfo{person}{Zhe Xu}, \bibinfo{person}{Zhixin Li},
  \bibinfo{person}{Qingwen Guan}, \bibinfo{person}{Dingshui Zhang},
  \bibinfo{person}{Qiang Li}, \bibinfo{person}{Junxiao Nan},
  \bibinfo{person}{Chunyang Liu}, \bibinfo{person}{Wei Bian}, {and}
  \bibinfo{person}{Jieping Ye}.} \bibinfo{year}{2018}\natexlab{}.
\newblock \showarticletitle{Large-Scale Order Dispatch in On-Demand
  Ride-Hailing Platforms: A Learning and Planning Approach}. In
  \bibinfo{booktitle}{\emph{Proceedings of the 24th ACM SIGKDD International
  Conference on Knowledge Discovery \& Data Mining}}. ACM,
  \bibinfo{pages}{905--913}.
\newblock


\bibitem[Yan et~al\mbox{.}(2020)]%
        {yan2020dynamic}
\bibfield{author}{\bibinfo{person}{Chiwei Yan}, \bibinfo{person}{Helin Zhu},
  \bibinfo{person}{Nikita Korolko}, {and} \bibinfo{person}{Dawn Woodard}.}
  \bibinfo{year}{2020}\natexlab{}.
\newblock \showarticletitle{Dynamic pricing and matching in ride-hailing
  platforms}.
\newblock \bibinfo{journal}{\emph{Naval Research Logistics (NRL)}}
  \bibinfo{volume}{67}, \bibinfo{number}{8} (\bibinfo{year}{2020}),
  \bibinfo{pages}{705--724}.
\newblock


\bibitem[Yao et~al\mbox{.}(2018)]%
        {yao2018deep}
\bibfield{author}{\bibinfo{person}{Huaxiu Yao}, \bibinfo{person}{Fei Wu},
  \bibinfo{person}{Jintao Ke}, \bibinfo{person}{Xianfeng Tang},
  \bibinfo{person}{Yitian Jia}, \bibinfo{person}{Siyu Lu},
  \bibinfo{person}{Pinghua Gong}, \bibinfo{person}{Jieping Ye}, {and}
  \bibinfo{person}{Zhenhui Li}.} \bibinfo{year}{2018}\natexlab{}.
\newblock \showarticletitle{Deep multi-view spatial-temporal network for taxi
  demand prediction}. In \bibinfo{booktitle}{\emph{Proceedings of the AAAI
  Conference on Artificial Intelligence}}, Vol.~\bibinfo{volume}{32}.
\newblock


\bibitem[Zhong et~al\mbox{.}(2020)]%
        {zhong2020queueing}
\bibfield{author}{\bibinfo{person}{Yueyang Zhong}, \bibinfo{person}{Zhixi Wan},
  {and} \bibinfo{person}{Zuo-Jun~Max Shen}.} \bibinfo{year}{2020}\natexlab{}.
\newblock \showarticletitle{Queueing Versus Surge Pricing Mechanism:
  Efficiency, Equity, and Consumer Welfare}.
\newblock \bibinfo{journal}{\emph{Equity, and Consumer Welfare (September 24,
  2020)}} (\bibinfo{year}{2020}).
\newblock


\bibitem[Zhou et~al\mbox{.}(2021)]%
        {zhou2021graph}
\bibfield{author}{\bibinfo{person}{Fan Zhou}, \bibinfo{person}{Shikai Luo},
  \bibinfo{person}{Xiaohu Qie}, \bibinfo{person}{Jieping Ye}, {and}
  \bibinfo{person}{Hongtu Zhu}.} \bibinfo{year}{2021}\natexlab{}.
\newblock \showarticletitle{Graph-Based Equilibrium Metrics for Dynamic
  Supply--Demand Systems With Applications to Ride-sourcing Platforms}.
\newblock \bibinfo{journal}{\emph{J. Amer. Statist. Assoc.}}
  \bibinfo{volume}{116}, \bibinfo{number}{536} (\bibinfo{year}{2021}),
  \bibinfo{pages}{1688--1699}.
\newblock


\bibitem[Zhou et~al\mbox{.}(2019)]%
        {zhou2019multi}
\bibfield{author}{\bibinfo{person}{Ming Zhou}, \bibinfo{person}{Jiarui Jin},
  \bibinfo{person}{Weinan Zhang}, \bibinfo{person}{Zhiwei Qin},
  \bibinfo{person}{Yan Jiao}, \bibinfo{person}{Chenxi Wang},
  \bibinfo{person}{Guobin Wu}, \bibinfo{person}{Yong Yu}, {and}
  \bibinfo{person}{Jieping Ye}.} \bibinfo{year}{2019}\natexlab{}.
\newblock \showarticletitle{Multi-Agent Reinforcement Learning for
  Order-dispatching via Order-Vehicle Distribution Matching}. In
  \bibinfo{booktitle}{\emph{Proceedings of the 28th ACM International
  Conference on Information and Knowledge Management}}.
  \bibinfo{pages}{2645--2653}.
\newblock


\end{thebibliography}

\newpage
\appendix

\section{Appendix}
\subsection{Proof of Proposition~\ref{prop:AdLessAs}}\label{proof:AdLessAs}

\begin{proof}
We start from the definition of $\Ad$ and consider the cases where $\mit\geq 0$ and $\mit<0$. As defined in Section \ref{sec:supply_view}, $M$ and $N$ are the total supply and demand volumes respectively. Then,
\begin{align}
    \nonumber\Ad &= \sum_{\ict}\nubarit\mit \\
        \nonumber &= \sum_{\mit\geq 0}\nubarit\mit + \sum_{\mit<0}\nubarit\mit \\
        &< \sum_{\mit\geq 0}\frac{\mutildeit}{N}\mit + \sum_{\mit<0}\frac{\mutildeit}{N}\mit \label{eq:proof_core}\\
        \nonumber &= \sum_{\ict}\frac{\mutildeit}{M}\mit\frac{M}{N} \\
        \nonumber &= \sum_{\ict}\mubarit\mit\cdot\frac{M}{N} \\
        &= \As\cdot\frac{M}{N} \label{eq:prop_result}
\end{align}
In \eqref{eq:proof_core}, we replace $\nuict$ with $\mutildeit$. When $\mit\geq 0$, we have $\mutildeit\geq\nuict$ by definition, and hence, $\sum_{\mit\geq 0}\nubarit\mit \leq \sum_{\mit\geq 0}\frac{\mutildeit}{N}\mit$, since $\nuict\geq 0, \mutildeit\geq 0, \forall (\ict)\in\calT$. When $\mit<0$, we have $\mutildeit < \nuict$, so $\sum_{\mit<0}\nubarit\mit < \sum_{\mit<0}\frac{\mutildeit}{N}\mit$.
\end{proof}

\subsection{Proof of Proposition~\ref{prop:decomp}}\label{proof:decomp}
\begin{proof}
Let $\vec v_T = (\alpha_{d, T}, \alpha_{s, T})$ and $\vec v_C = (\alpha_{d, C}, \alpha_{s, C})$. The shift vector is $\vec v_{\text{shift}} = \vec v_T - \vec v_C$.  Then
\[\tautotal = \|\vshift\|_2.\]
Define the unit vectors
\[b_{\text{vol}} = (\frac{1}{\sqrt{2}}, \frac{1}{\sqrt{2}}), \qquad \vec b_{\text{dist}} = (\frac{1}{\sqrt{2}}, -\frac{1}{\sqrt{2}}),\]
which are the unit vectors in the direction of the pink and blue lines from Figure~\ref{fig:global_frontier}, respectively.

Now consider projecting $\vshift$ onto the line spanned by $b_{\text{vol}}$. This projection is
\[\proj_{\vec b_{\text{vol}}} \vshift = \tauglobal \vec b_{\text{vol}},\]
where $\tauglobal$ is defined to be the coefficient and is given by
\[\tauglobal = \frac{\vshift \cdot \vec b_{\text{vol}}}{\|\vec b_{\text{vol}}\|^2} = \frac{1}{\sqrt{2}} \left[ \alpha_{s,T} + \alpha_{d,T} - \alpha_{s,C} - \alpha_{d,C}\right]\]

For $b_{\text{dist}}$, we have
\[\proj_{\vec b_{\text{dist}}} \vshift = \tauglobal \vec b_{\text{dist}},\]
where $\taucond$ is defined to be the coefficient and is given by
\[\taucond = \frac{\vshift \cdot \vec b_{\text{dist}}}{\|\vec b_{\text{dist}}\|^2} = \frac{1}{\sqrt{2}} \left[ \alpha_{s,T} - \alpha_{s,C} - \alpha_{d,T} + \alpha_{d,C}\right]\]

It can be verified that
\begin{align*}
\tauglobal^2 &= \frac{1}{2}\left[(\alpha_{d, T} - \alpha_{d, C})^2 + (\alpha_{s, T} - \alpha_{s, C})^2 + 2(\alpha_{d, T} - \alpha_{d, C})(\alpha_{s, T} - \alpha_{s, C}) \right]\\
\taucond^2 &= \frac{1}{2}\left[(\alpha_{d, T} - \alpha_{d, C})^2 + (\alpha_{s, T} - \alpha_{s, C})^2 - 2(\alpha_{d, T} - \alpha_{d, C})(\alpha_{s, T} - \alpha_{s, C}) \right]
\end{align*}
and thus the decomposition $\tautotal^2 = \tauglobal^2 + \taucond^2$ holds.

\end{proof}

\end{document}